\documentclass{emulateapj}
\usepackage{amsmath}
\usepackage{color}
\usepackage{enumerate}

\newcommand{\Msun}{M_\odot}
\newcommand{\sci}[1]{\times10^{#1}}
\newcommand{\fig}[1]{Figure \ref{fig:#1}}
\newcommand{\figs}[2]{Figures \ref{fig:#1} and \ref{fig:#2}}

\newcommand{\eqn}[1]{Equation (\ref{eqn:#1})}

\newcommand{\lcdm}{$\Lambda$CDM}
\newcommand{\kms}{km\,s$^{-1}$}

\newcommand{\degrees}{^\circ}

\newcommand{\rev}{}

\shorttitle{Tidal debris dispersal in MW-sized halo}
\shortauthors{Ngan, Carlberg, Bozek, et al}

\begin{document}


\title{{Dispersal} of tidal debris in a {Milky-Way-sized} dark matter halo}



\author{Wayne Ngan$^1$, Raymond G. Carlberg$^1$, Brandon Bozek$^{2,3}$,
Rosemary F. G. Wyse$^3$, Alexander S. Szalay$^3$, Piero Madau$^4$}

\affil{$^1$ Department of Astronomy and Astrophysics, University of Toronto, Toronto ON, M5S 3H4, Canada \\
\rev{$^2$ Department of Astronomy, The University of Texas at Austin, Austin TX, 78712, USA} \\
$^3$ Department of Physics and Astronomy, The Johns Hopkins University, Baltimore MD, 21218, USA \\
$^4$ Department of Astronomy and Astrophysics, University of California, Santa Cruz CA, 95064, USA}
\email{ngan@astro.utoronto.ca}




\begin{abstract}
We simulate the tidal disruption of a collisionless N-body globular star cluster in \rev{a total of 300} different orbits 
\rev{selected} to have \rev{galactocentric} radii between 10 and 30 kpc in four dark matter halos:
(a) {a} spherical halo with no subhalos, (b) {a} spherical halo with subhalos, (c) {a} realistic halo with no
subhalos, and (d) {a} realistic halo with subhalos. This allows us to isolate and study how the 
halo's {(lack of) dynamical symmetry} and substructures affect the {dispersal} of tidal debris. The realistic
halos are constructed from the snapshot of the Via Lactea II simulation at redshift zero. We find
that the {overall} halo's {lack of dynamical symmetry} \rev{disperses tidal debris to make the streams fluffier,}
consistent with previous studies of tidal debris of dwarf galaxies in larger orbits {than ours in this study}. 
On the other hand, subhalos in realistic potentials {can
locally enhance the densities along streams}, making streams denser than their counterparts
in smooth {potentials}. \rev{We show that many long and thin streams can survive in a realistic and lumpy halo
for a Hubble time.} This suggests that upcoming stellar surveys will likely uncover more thin streams which 
may contain density gaps that have been shown to be promising probes for dark matter substructures.

\end{abstract}

\section{Introduction}
\label{sec:introduction}

\rev{Recent surveys have discovered dynamically cold tidal debris of many disrupted stellar
systems---likely globular clusters or low mass dwarf galaxies---in the inner tens
of kiloparsecs of the Milky Way
\citep{grillmair2010,bonaca2012,grillmair2014,koposov2014,martin2014,bernard2014}.
These debris, arranged in thin streams of stars, are powerful probes for the 
structure of our own galaxy. In particular, cold streams
are more useful than hot streams for constraining the shape of the Milky Way's halo 
\citep[see][and references therein]{lux2013}. Also, because of their low velocity dispersion,
cold streams exhibit density variations, or ``gaps'', that are caused by dark matter subhalos
\citep{yoon11, carlberg12, pal5gaps, carlberg13, gd1, NC14, N14, erkal2015}.}



Many theoretical studies of tidal streams rely on idealized profiles, such as logarithmic
or Navarro-Frenk-White profiles \citep{nfw}, as the dark matter halo potentials of the host galaxies. Although
idealized profiles have analytic forms and are simple to compute, {they result from spherical averaging
of many simulations of dark matter halos \citep{nfw, navarro04,navarro10}, and the averaging} does not capture the realistic
details inside the
individual potentials. Furthermore, both theoretical \citep{dc91, js02, zemp2009} and observational
\rev{\citep{law09,lm2010,dw2013,vh2013}} efforts concluded that Milky Way's
dark matter halo is not spherical but triaxial. Some studies model the shapes of halos by 
introducing triaxial parameters into the radial coordinate of the idealized profiles 
\citep[e.g.,][]{lm2010}. In this study, however, we take an even more general approach
\rev{to simulate tidal streams directly in the potential of Via Lactea II (VL-2), a high-resolution
dark matter halo simulation in the \lcdm\ cosmological context \citep{diemand08},} without
\rev{fitting the halo} to any idealized profiles.

\rev{Tidal stream simulations in the VL-2 potential was first done by \citet{bonaca2014}, where they gave VL-2
its original N-body treatment, but generated tidal streams using the Streakline method \citep{streakline}.}
In \citet{N14} we took a different approach, where we constructed a potential model using one snapshot of VL-2,
but treated the tidal streams as N-body problems.
The self-consistent field method \citep{scf} allowed us to construct accurate potential
models \rev{that} are optimized for dark matter halos. We furthermore isolated the subhalos found
in the VL-2 halo, so we were able to construct two models: a ``smooth'' VL-2 halo with no
subhalos, and a ``lumpy'' VL-2 halo with the subhalos originally found. These two models can be
used to study the effects of subhalos on tidal streams \rev{in order to} shed light
on the ``missing satellites problem'' \rev{\citep{klypin99, moore99, yoon11, NC14, N14}},
\rev{and we continue to use our models for this study.}

Recently, \citet{bonaca2014,N14,pearson2014,fardal15} showed that streams inside \rev{non-spherical} potentials
exhibit fluffy features \rev{dubbed ``stream-fanning''} which are not found
in streams inside spherical potentials. \rev{This effect has been invoked by \citet{sesar16} to explain
the puzzling orbit and length of the recently observed Ophiuchus stream \citep{bernard2014,sesar2015}.}
\rev{This suggests that spherical potentials are no longer suitable for modeling tidal streams.
It is important to note that while stream-fanning implies a non-spherical potential, the converse is not true;
not all streams inside non-spherical potentials exhibit stream-fanning.} The results of our study here, as we will {quantify},
indicate that while streams inside a realistic potential
are {in general wider and more diffuse} than streams inside a spherical halo,
\rev{thin streams can still exist even under the influence of a realistic halo and subhalos.}

The goal of this study is to investigate how dispersed tidal streams are in a realistic halo, and
\rev{whether we should be optimistic that large stellar surveys in the near future will be able to
uncover more cold and thin streams similar to Pal-5 \citep{odenkirchen01} or GD-1 \citep{grillmair2006}.}
We simulate a total of 300 tidal streams as {self-gravitating} N-body simulations
of the same globular star cluster orbiting inside \rev{the following 
cases of host halos}: (a) {a} spherical halo with no subhalos, (b) {a} spherical halo with subhalos,
(c) {the} VL-2 halo with no subhalos, and (d) {the} VL-2 halo with subhalos. In all cases, the host's and subhalos'
potentials are not evolving in time, although subhalos orbit around the host's potential as test masses.
\rev{\citet{bonaca2014} fitted the VL-2's potential at different times to triaxial idealized profiles while allowing each 
parameter to vary radially, and they found that each parameter changed by no more than $\lesssim10\%$ over the 
last 6 Gyr. Using} a time dependent potential, \rev{especially at the early stages of the halos' formation,}
is an important step in the future, but we believe that the effects of a realistic potential itself
is worth studying before including the additional complication of time dependence.

This paper is organized as follows. Section \ref{sec:method} is the method section which contains several
subsections. Section \ref{sec:method_vl2} summarizes the SCF method to obtain the potential
from the VL-2 halo and subhalos. Section \ref{sec:method_spherical} describes our spherical halo
which will be compared against the VL-2 halo. Section \ref{sec:subhalo_finding} justifies the mass
range of subhalos included in the lumpy halo cases. Section \ref{sec:method_progenitor} describes
the initialization of the stream progenitor, including its density profile and its orbits.
Section \ref{sec:results} contains our results. {In particular, Sections \ref{sec:effects_of_the_underlying_halo}
and \ref{sec:dispersion_of_tidal_debris} compare the dispersal of tidal debris in the spherical
and the VL-2 halos, and \rev{Sections \ref{sec:effects_of_subhalos} and \ref{sec:the_densest_streams_in_the_vl2_halos} 
discuss the effects of subhalos and how they enhance instead of lower the densities of streams.}
Section \ref{sec:conclusion} summarizes our results and points to future work.


\section{Method}
\label{sec:method}

\subsection{VL-2 Halo and Subhalo Potentials}
\label{sec:method_vl2}

The smooth and lumpy VL-2 potentials are obtained
using the same \rev{code} detailed in \citet{N14}, which uses a combination of  the self-consistent
field (SCF) method \citep{scf} and a halo finder code
to construct the gravitational potentials inside the host halo and subhalos.
\rev{The N-body simulations of the streams are run using the public
version\footnote{\url{http://www.mpa-garching.mpg.de/gadget/}}
of {\sc Gadget-2} \citep{gadget}. The accelerations due to both the host halo and subhalos are added to the
stream particles after their N-body forces have been computed. We impose a maximum time step of 1 Myr and softening
of 5 pc in each particle.}

The SCF method has also been applied by \citet{hex} for the dark matter halos in 
the Aquarius simulations \citep{aquarius}. In this section we briefly summarize the method,
\rev{which} begins by specifying a set of basis functions $\Phi_{nlm}(\mathbf{r})$ and solves the Poisson
equation in spherical coordinates $\mathbf{r}\equiv(r,\theta,\phi)$ for the potential in the form
\begin{equation}
	\Phi(\mathbf{r}) = \sum_{n=0}^{n_{max}} \sum_{l=0}^{l_{max}} \sum_{m=0}^l A_{nlm} \Phi_{nlm}(\mathbf{r}).
	\label{eqn:phinlm}
\end{equation}
Given a list of particle positions and masses stored in an existing snapshot of a  \rev{dark matter halo}
simulation, the SCF method provides a recipe to
obtain the basis coefficients $A_{nlm}$, and then the force field $-\nabla \Phi$ can be computed analytically 
by differentiating \eqn{phinlm}. In this study, following \citet{hex} and \citet{N14}, the zeroth
order basis function is in the form of a Hernquist profile \citep{hernquist}, and the higher
order radial and angular deviations are polynomials in $r$ of degree $n$ and spherical harmonics in
($\theta$, $\phi$) of orders $l$ and $m$.

The smooth VL-2 potential is obtained by the SCF method using order 10
(where $n_{max}=l_{max}=\mathrm{order}$) on the VL-2 main halo after its subhalo particles have been removed. 
This order ensures that the overall shape of the main halo is captured without the lumpiness
due to either the presence of subhalos or the voids after the subhalos are removed, since a \rev{polynomial} of degree 10
is not sufficient to model more than 10,000 subhalos. \rev{As documented in \cite{N14}, the
decomposition of the main halo is performed inside the Virial radius (as reported by the halo finder),
which is 400 kpc. When using an order 10 polynomial to decompose the main halo, we are only capturing features of
$\sim40$ kpc in size. The scale radius of the largest subhalo in VL-2 is 6.4 kpc. Therefore, an order 10 polynomial
for the main halo would not be sensitive to the either the presence or absence of subhalos.}
The details for subhalo abundance \rev{are} discussed in Section \ref{sec:subhalo_finding}.

\subsection{Spherical Halo Potential}
\label{sec:method_spherical}

The spherical halo is modeled using a Navarro-Frenk-White profile \citep[NFW;][]{nfw} whose enclosed mass
profile is
\begin{equation}
	M(r) = \frac{v_h^2 r_h}{G} \left[ \ln\left(1+\frac{r}{r_h}\right)-\frac{r/r_h}{1+r/r_h} \right]
\end{equation}
where $r_h$ and $v_h$ are the scale parameters.
\rev{We perform a least-squares fit to the VL-2 halo's mass profile at redshift zero using the 
NFW profile as a model. Because our streams in this study are restricted inside $10\,\mathrm{kpc}<r<30\,\mathrm{kpc}$ of
the halo, as we show in Section \ref{sec:method_progenitor}, the fit is performed for 
that region only. Our best fit $r_h$ and $v_h$ are 19.1 kpc and 421 \kms. This gives circular velocities of
162 and 194 \kms at $r=10$ and 30 kpc, respectively.}




\rev{\fig{halo_masses} shows the ratio of the enclosed mass profiles for our VL-2 to spherical halos.
Between 10 and 30 kpc, the mass profiles agree to within a few percent. However, at less than 
10 kpc the mass profiles deviate by almost 50\%. This is not surprising, as it has already been shown 
that VL-2's profile has an inner slope of $\gamma=1.24$ \citep{diemand08} versus $\gamma=1$ in an 
NFW profile. Therefore, it is impossible to fit both the inner and the outer parts of the halo simultaneously.
Despite this, we use the NFW profile for the spherical halo because of its simplicity and popularity in the
literature. For the rest of our study, we focus on streams which only attain orbits between 10 to 30 kpc
in order to facilitate a fair comparison between VL-2 and spherical halos.
This is discussed in detail in Section \ref{sec:method_progenitor}.}

\rev{The matching halo mass profiles, at least for $10\,\mathrm{kpc}<r<30$ kpc, ensures that} the dimensionless tidal scales are
comparable between the two types of halos for the same star cluster mass and similar orbits. 
The {dimensionless} tidal scale {$s$, which is the ratio between the star cluster's tidal radius and
its orbital perigalacticon radius, can also be written as}
\begin{equation}
	s\equiv\left( \frac{m}{M(r)} \right)^{1/3},
	\label{eqn:tidal_scale}
\end{equation}
where $m$ is the star cluster mass, and $M(r)$ is the mass enclosed inside radius $r$ of the host halo.
\citet{johnston1998} and \citet{johnston2001} found that in a spherical potential, $s$ determines
the {spreads in} energy and angular momentum, which in turn determine the width and length of a stream.
\rev{Our goal is to study the difference between two identically distributed ensembles of streams in 
the VL-2 halo and a spherical halo with similar tidal scales}.


\begin{figure}
	\includegraphics[width=3.4in]{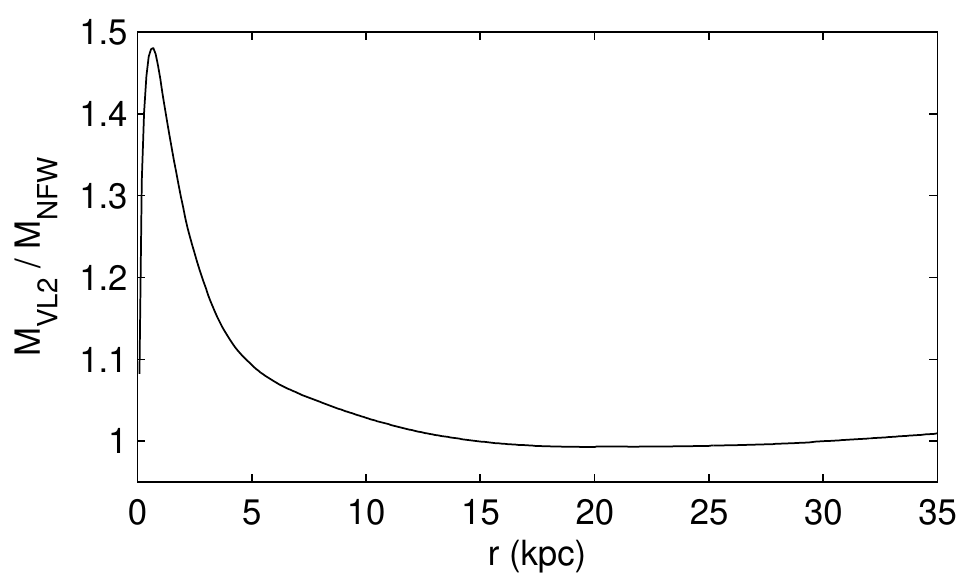}
	\caption{Enclosed masses of the two smooth halos used in this study.
	The mass of the VL-2 is computed by directly summing the masses of all the
	particles inside $r$, and the mass of the spherical halo is obtained analytically from the NFW profile.
	Their masses agree to within a few percent for $10\,\mathrm{kpc} < r < 30\,\mathrm{kpc}$ kpc, where our
	streams are \rev{selected} to orbit in.}
	\label{fig:halo_masses}
\end{figure}

\subsection{Subhalo Finding}
\label{sec:subhalo_finding}

The lumpy host {halos} are simply the smooth host {halos} 
with extra subhalos orbiting in them. The subhalos of the host halos are identified and
extracted from the zero redshift snapshot of VL-2
\rev{using the {\sc Amiga Halo Finder\footnote{
\url{http://popia.ft.uam.es/AHF/Download.html}}}. Inside the main halo of VL-2, {we consider only
the subhalos which are the immediate satellites of the main halo, such that all the satellites
of a given subhalo (including all the hierarchically smaller satellites) are all considered as
part of that subhalo.}}

For each subhalo, the halo finder returns the positions and velocities
which are used to initialize the subhalo's orbit around the main halo, as well as masses and scale radii which are
used to construct the subhalo's potential using the SCF method \rev{as described above, but} with $n_{max}=4$
and $l_{max}=0$. To ensure a fair comparison,
all the subhalos in both the VL-2 and spherical host halos are modeled identically\rev{, and their
orbits are initialized identically as well}. 
The subhalos are assumed not to interact with each other and orbit the host halos as test masses,
{and they} use the positions and velocities \rev{at redshift zero of VL-2} as orbital initial conditions.

We identify 11,523 subhalos in total \rev{whose masses range from $1.2\sci{6}\Msun$ to $4.7\sci{9}\Msun$} 
in the main halo in the zero redshift snapshot of VL-2. This is consistent with the abundance reported by 
\citet{diemand08}. In each lumpy case, we further consider two sub-cases with two different
mass ranges $M_{sub}$ of subhalos: (i) all subhalos with $M_{sub}>10^7\Msun$ and (ii) all subhalos with
$M_{sub}>5\sci{6}\Msun$. Our halo finder reports totals of 1087 and 2007 subhalos for the two mass ranges,
respectively. \rev{These lower limits in our mass ranges are chosen such that their contributions to the dispersal
of tidal debris become significant. As we show in our results later, expanding the range of
subhalos from (i) to (ii) has a negligible effect on tidal debris; meanwhile, \citet{NC14} showed that the inclusion of 
$M_{sub}<10^8\Msun$ subhalos can open up large gaps, which may significantly affect the distribution of
debris along the stream.} As opposed to \citet{NC14} and \citet{N14}, here we do not impose an upper limit for subhalo masses
because our goal is to investigate the \rev{global} perturbations on tidal streams \rev{where
many stars in the stream are affected (ie. how dispersed the debris is), \rev{rather than local perturbations
which only affect small parts (ie. gaps) of the stream.}
Note that this means the subhalo with the highest mass in this study is more massive than those in
\citet{NC14} and \citet{N14} which were limited to $M_{sub}<10^8\Msun$}. 



Similar to \citet{NC14} and \citet{N14}, in our actual simulations we do not include the subhalos
which interact minimally with the streams, as some subhalos have pericentric distances {that} are
greater than the streams' apocentric distances. As explained below, our stream orbits are restricted
to a maximum apocentric distance of 30 kpc. Therefore, subhalos with pericentric distances greater than 40 kpc can be
safely eliminated. For the mass ranges of subhalos mentioned above, only 381 and 674 subhalos, respectively, remain for
the VL-2 potential, and \rev{306} and \rev{547} subhalos, respectively, remain for the spherical potential.
\rev{The subhalos in both the VL-2 and spherical halos spend medians of $\sim0.4$ Gyr in total (out of 10 Gyr which is 
the duration of each stream simulation) in the inner 40 kpc of their respective halos.
Since we include more subhalos in the VL-2 halo than in the spherical halo, subhalos are more
likely to impact our streams in the former case. However, this does not affect our conclusions at the end.}

Because the lumpy halos are essentially the smooth halos with extra subhalos, lumpy halos are more massive
than smooth halos. In the VL-2 halo, 381 and 674 subhalos are about
1.6\% and 1.7\% of the mass of the smooth halo \rev{enclosed in its Virial radius 400 kpc}.
In the spherical halo, 306 and 547
subhalos are about 1.2\% and 1.3\% of the \rev{total} mass of the smooth halo \rev{truncated at 400 kpc}. 
In \citet{NC14} and \citet{N14}, stream progenitors with the same initial conditions travel in roughly
the same orbit in both smooth and
lumpy halos because each subhalo's mass is low enough that each subhalo individually does not affect
the streams' orbits much. In this study, however, stream progenitors travel in very different orbits
in smooth and lumpy potentials because most of the subhalo masses are concentrated in the most massive
subhalos, which can affect the streams' orbits. {As shown in \citet{yoon11}, the characteristic energy
that a $\sim10^9\Msun$ subhalo deposits into a stream is $\sim10^4\, \mathrm{km}^2\mathrm{s}^{-2}$,
and this corresponds to the orbital energy $E\propto v_{orbit}^2 \sim (200\, \mathrm{km\,s}^{-1})^2$
of a GD-1-like stream whose orbit oscillates radially between $15-30$ kpc.
Therefore, subhalos more massive than $\sim10^9\Msun$ can significantly affect the orbit of the
stream progenitor, hence the entire stream itself.} This is the reason the smooth and lumpy halos cannot
be compared directly using \rev{individual} streams in each case when simulations include high mass subhalos. Smooth and
lumpy halos can only be compared by the statistics of \rev{ensembles of} streams as we discuss in our results in
Section \ref{sec:results}.


\subsection{Stream Progenitor and Orbits}
\label{sec:method_progenitor}


The progenitor for all streams in this study is \rev{the same} {self-gravitating N-body}
star cluster \rev{of $N=50000$ particles} following a King profile with $w=2$, where $w$ is the ratio
between the central potential
and velocity dispersion of the cluster. We initialize the cluster with core radius $r_0=0.05$ kpc
and mass $m=5\sci{4}\Msun$, and this yields a cluster with tidal radius $\sim0.15$ kpc
(beyond which the density is zero) and velocity dispersion $\sim1$ \kms. \rev{This star cluster
is typically dissolved in a few gigayears and produces a thin stream similar to GD-1 in our
orbits described below.}


The random orbits in our simulations are selected as follows. We first place 10,000 random points following
a uniform distribution inside a spherical shell between $15$ and $30$ kpc in radius. Each point is also
assigned uniformly random velocities between $-300$ and $300$ \kms\ for each of the $v_x,v_y,v_z$ components.
We then use these positions and velocities as initial conditions to integrate 10,000 test particle orbits
for 10 Gyr inside both smooth VL-2 and spherical potentials.

In the VL-2 potential, we randomly select 50 orbits such that their
galactocentric distances $r$ are bounded by $10\,\mathrm{kpc}<r<30\,\mathrm{kpc}$ for 10 Gyr.
The resulting eccentricity distribution of these 50 orbits selected from this process is shown in
\fig{eccentricities}. Here eccentricity is defined as
\begin{equation}
	e = \frac{r_{max}-r_{min}}{r_{max}+r_{min}}
\end{equation}
where $r_{max}$ and $r_{min}$ are the maximum and minimum $r$ attained by the test particle
in 10 Gyr. The initial conditions
of these 50 orbits serve as the stream progenitor's initial conditions for the streams' orbits inside both 
 smooth and lumpy VL-2 potentials.

In the spherical potential, we repeat a similar process which also selects 50 orbits with
the same limiting $r$. However, to ensure a fair comparison, the orbits are selected such that they produce identical
eccentricity distribution, shown in \fig{eccentricities}, as the VL-2 case. Similarly, these 50 \rev{orbital
initial conditions are applied to the stream progenitors} for the streams inside both the smooth and lumpy spherical potentials.


Note that a star cluster with a density profile \rev{and orbits} described above is \rev{typically dissolved} after $\sim5$
Gyr, as shown in \fig{massloss}.
This ensures that computational efforts are well spent, especially because each individual stream is a fully
self-gravitating simulation. Our goal is to study the dynamical evolution of tidal tails and
not the progenitors, so N-body particles that remain bound to the progenitors are irrelevant to our study.
We have repeated our simulations with a more tightly bound progenitor with the same mass and tidal radius.
We find that although the resulting streams are represented with {fewer} particles, our results are almost
identical to the ones produced using the progenitor described above, which we will {adopt} for the rest
of our results.

Also demonstrated in \fig{massloss} is that, for the same type of host halo and orbits, the median mass loss in
smooth and lumpy halos are almost the same. \rev{In the VL-2 halo, subhalos can help dissolve the progenitor in
the outlier cases, but typically this effect is very small.} This means that mass loss due to subhalo shocking
\rev{on the progenitor} is rare, and that the \rev{orbits of the particles that are bound to the progenitor should be irrelevant.
Therefore, rather than treating the progenitor and the stream as an N-body system, ``shortcut'' methods
\citep[e.g.][]{streakline,bonaca2014,gibbons14,fardal15} which eject test particles at the progenitor's Lagrange points
are promising alternatives. These methods potentially allow realistic streams to be generated quickly,
so they are worth adopting in the future.}

\begin{figure}
	\includegraphics[width=3.5in]{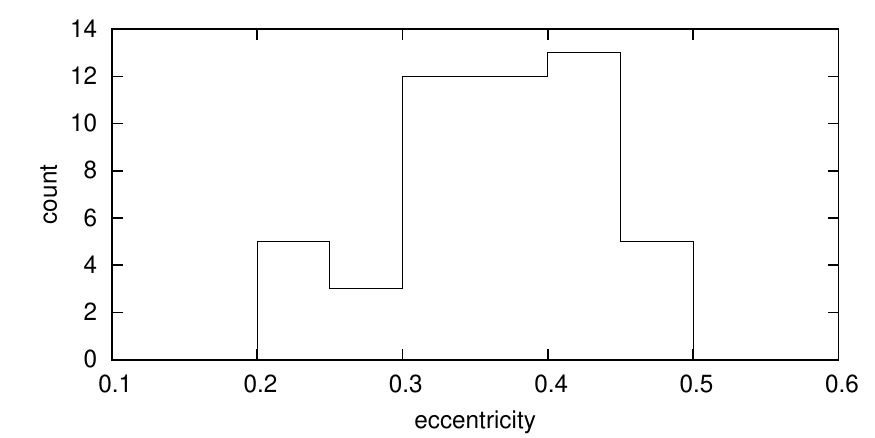}
	\caption{Eccentricities of the random orbits of the 50 streams \rev{simulated} in each halo potential.
	\rev{These orbits are all selected such that their galactocentric distances $r$ are bounded by 
	$10\,\mathrm{kpc}<r<30\,\mathrm{kpc}$ for 10 Gyr without subhalos. This range is similar to
	the orbits of observed streams such as Pal-5 and GD-1 which are thin and rich in substructures.}
	The sets of orbits in the VL-2 and spherical potentials are
	different, but they have identical eccentricity distribution.}
	\label{fig:eccentricities}
\end{figure}

\begin{figure*}
	\includegraphics[width=7in]{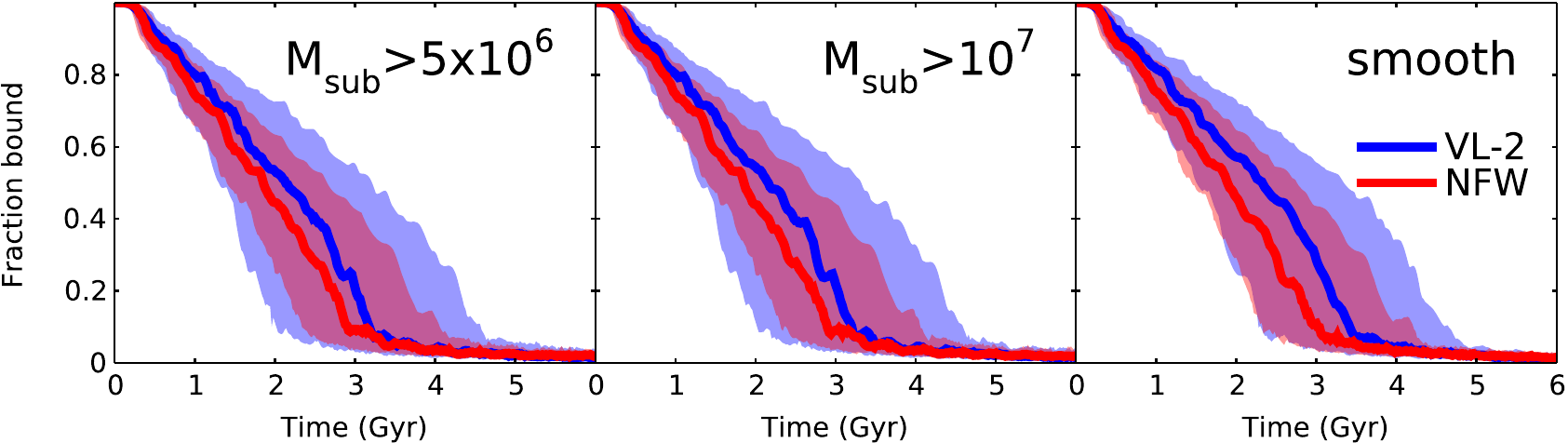}
	\caption{Fractions of particles enclosed in a radius 0.16 kpc of the progenitors as functions of time.
	Each solid line represents the median of 50 streams at each instant of time, \rev{and the colored areas enclose
	the points between the 15th and the 85th percentiles ($\sim1 \sigma$) of the distribution at each instant of time.
	Each panel shows the mass range of subhalos $M_{sub}$ present in the simulations,
	as labelled in units of solar masses. The similarity between these panels shows that the mass loss is primarily
	caused by the tidal stripping at the progenitors'} pericentric approaches, and not by subhalo shocking.}
	\label{fig:massloss}
\end{figure*}

%
%

\section{Results}
\label{sec:results}

\subsection{Effects of the Underlying Halo}
\label{sec:effects_of_the_underlying_halo}

\begin{figure*}[h!]
	\centering
	\Large{Spherical NFW halo}\\
	\smallskip
	\includegraphics[width=3.4in]{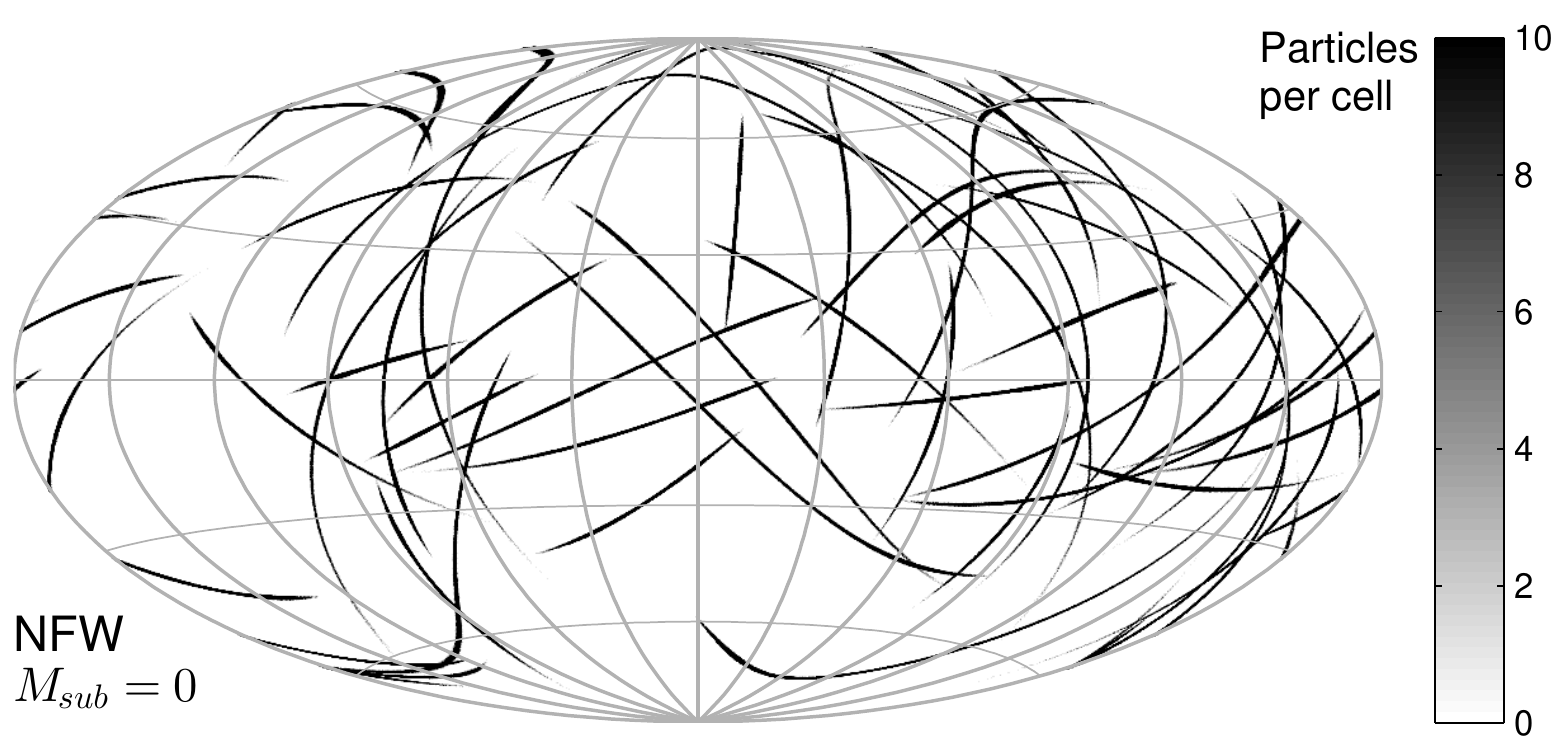}
	\hspace{0.05in}
	\includegraphics[width=3.4in]{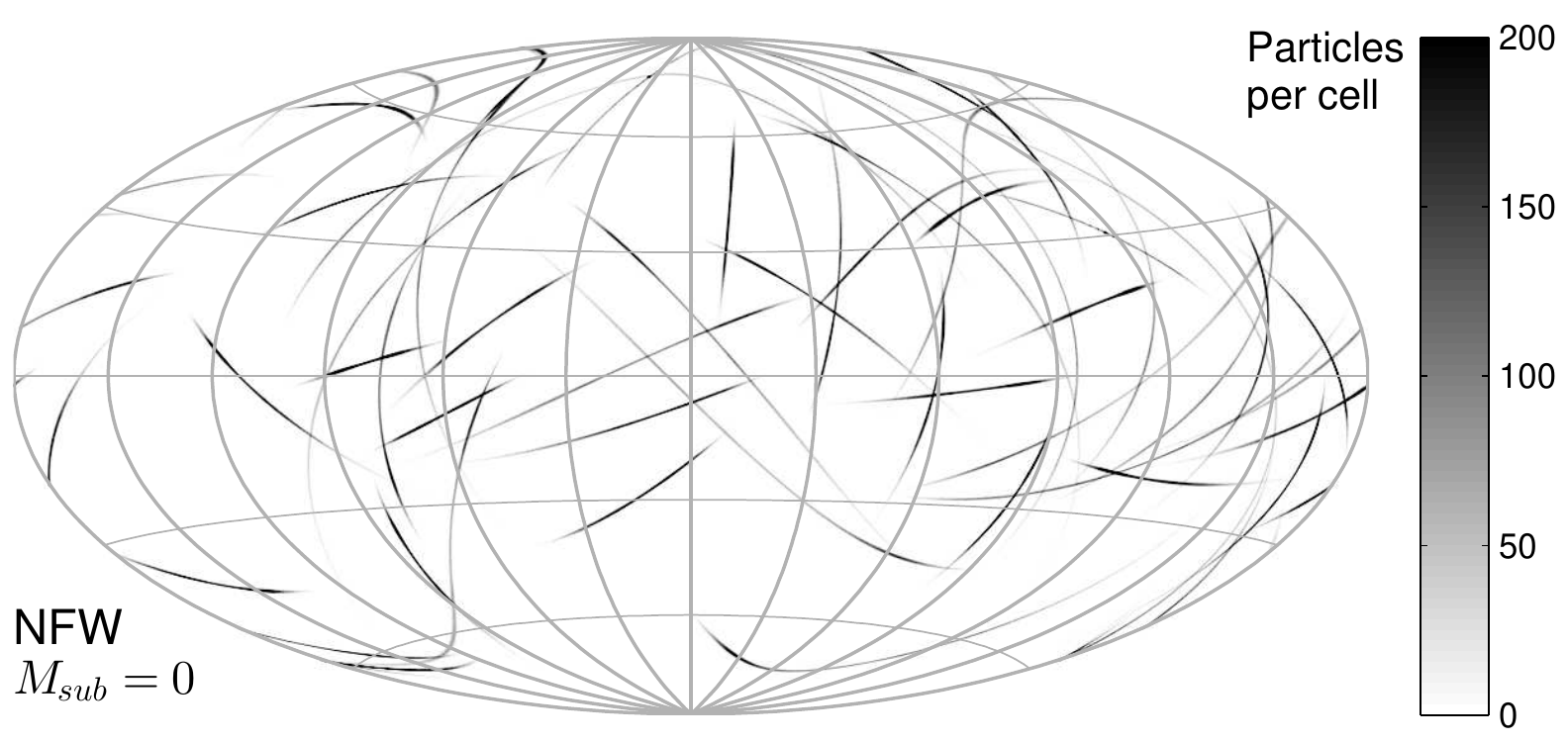} \\
	\includegraphics[width=3.4in]{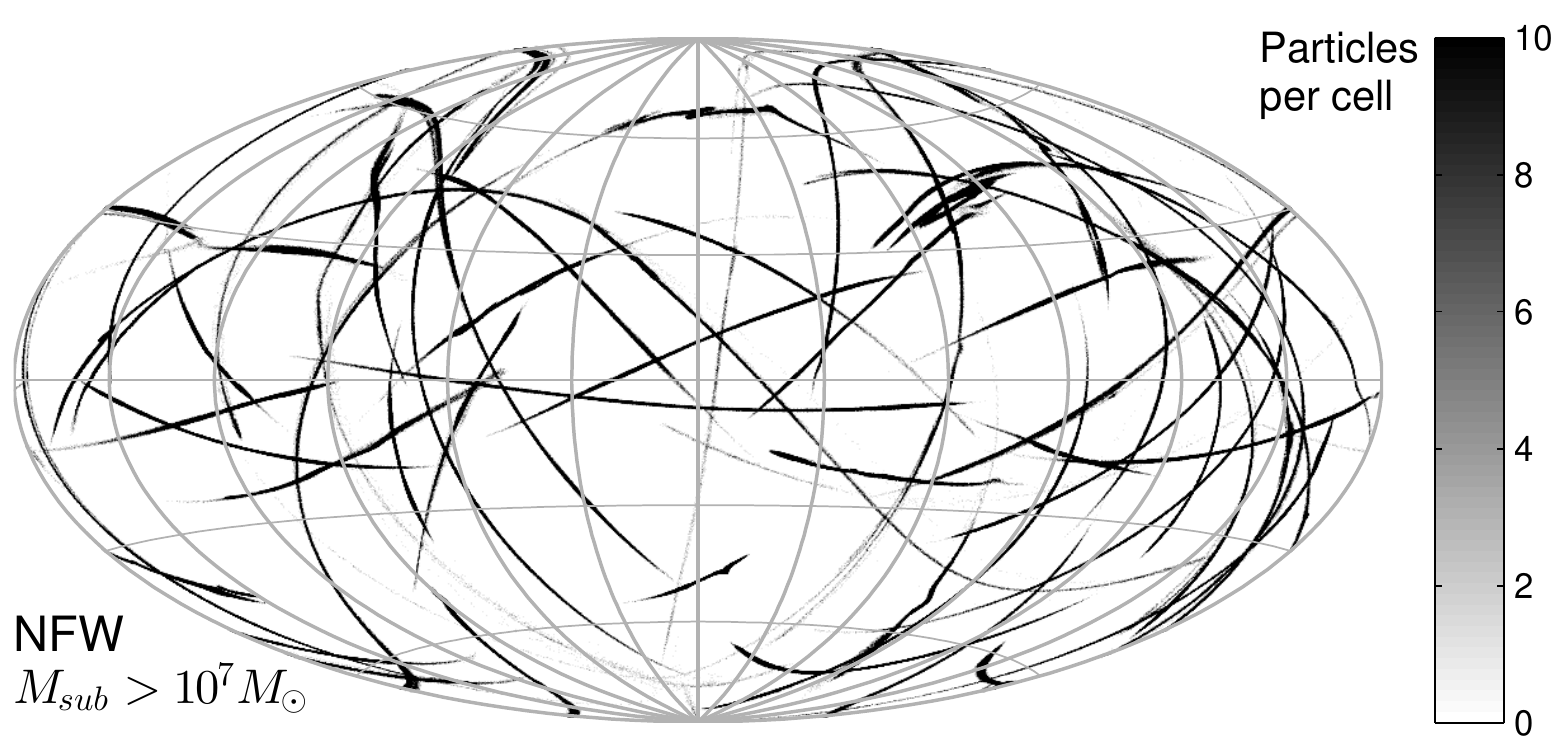}
	\hspace{0.05in}
	\includegraphics[width=3.4in]{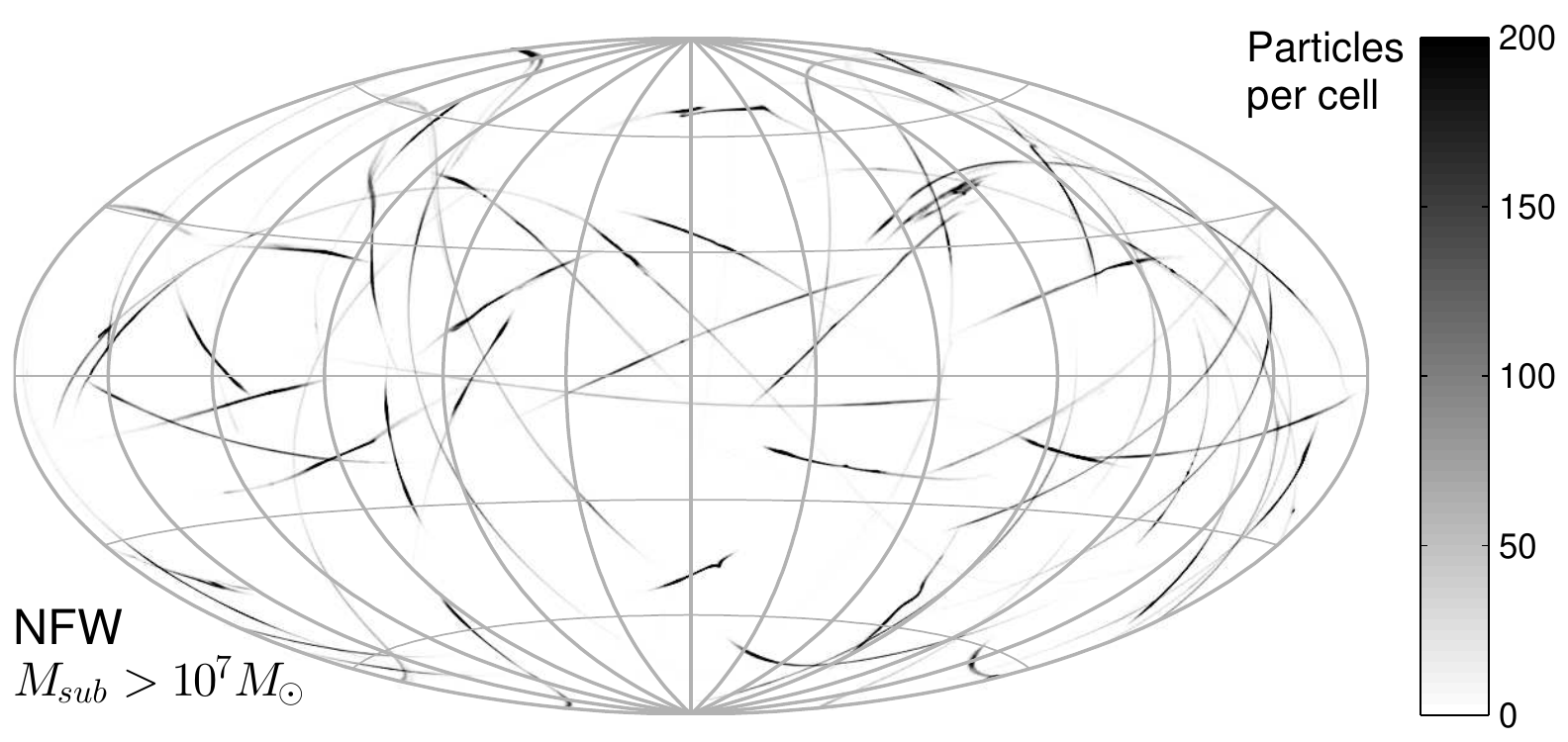}
	\caption{Hammer-Aitoff equal-area projections of 50 streams on the sky at 6 Gyr as seen from
	the galactic center. Each panel
	shows the combined surface density of particles in $0.3\degrees$ cells. Top and
	bottom rows show the streams in a smooth spherical halo and a lumpy spherical halo with 
	$M_{sub}>10^7\Msun$, respectively. Left and right columns show the same maps, but their gray 
	scales are adjusted to {emphasize} the diffuse and dense cells, respectively.
	All orbits of the stream progenitors are selected to have their galactocentric distances $r$ bounded by
	$10\,\mathrm{kpc}<r<30\,\mathrm{kpc}$ (see text for details). The orbital initial conditions of the streams
	are the same in 	top and bottom	panels, but the streams travel in different orbits due to the influence
	of the subhalos. The case where $M_{sub}>5\sci{6}\Msun$, not shown here, produces almost identical maps to the bottom panels.}
	\label{fig:sky_NFW}
\end{figure*}

\begin{figure*}
	\centering
	\Large{Via Lactea II halo}\\
	\bigskip
	\includegraphics[width=3.4in]{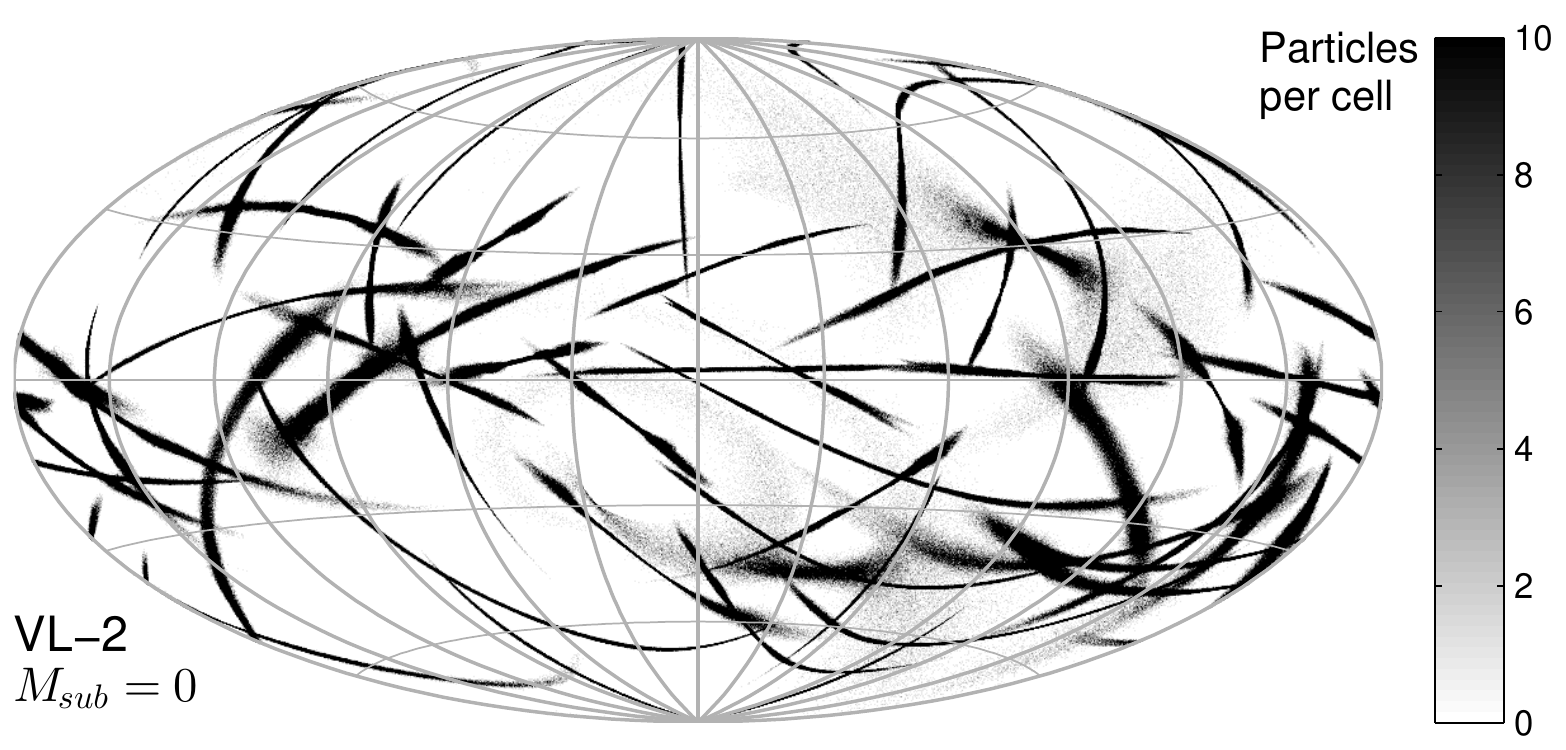}
	\hspace{0.05in}
	\includegraphics[width=3.4in]{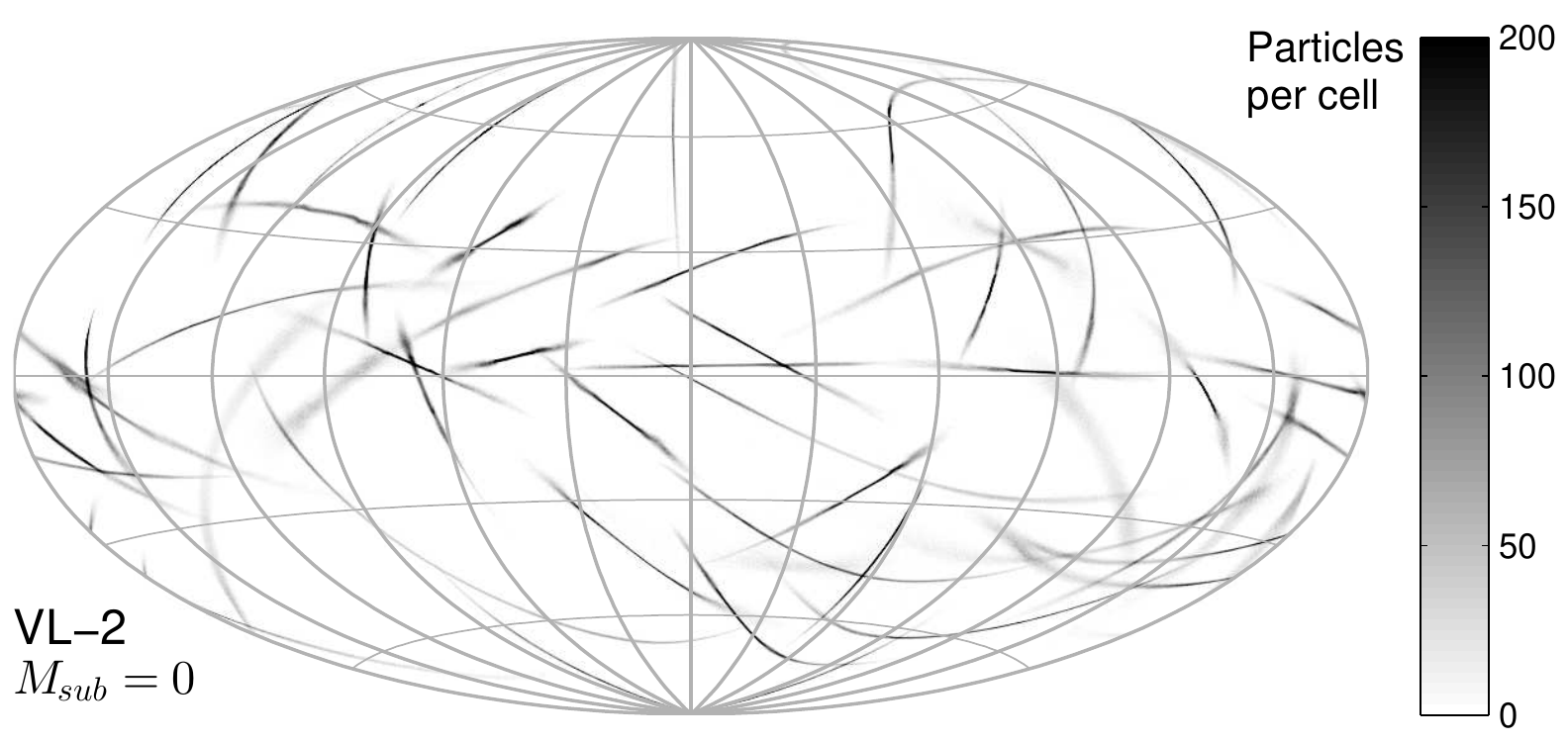} \\
	\includegraphics[width=3.4in]{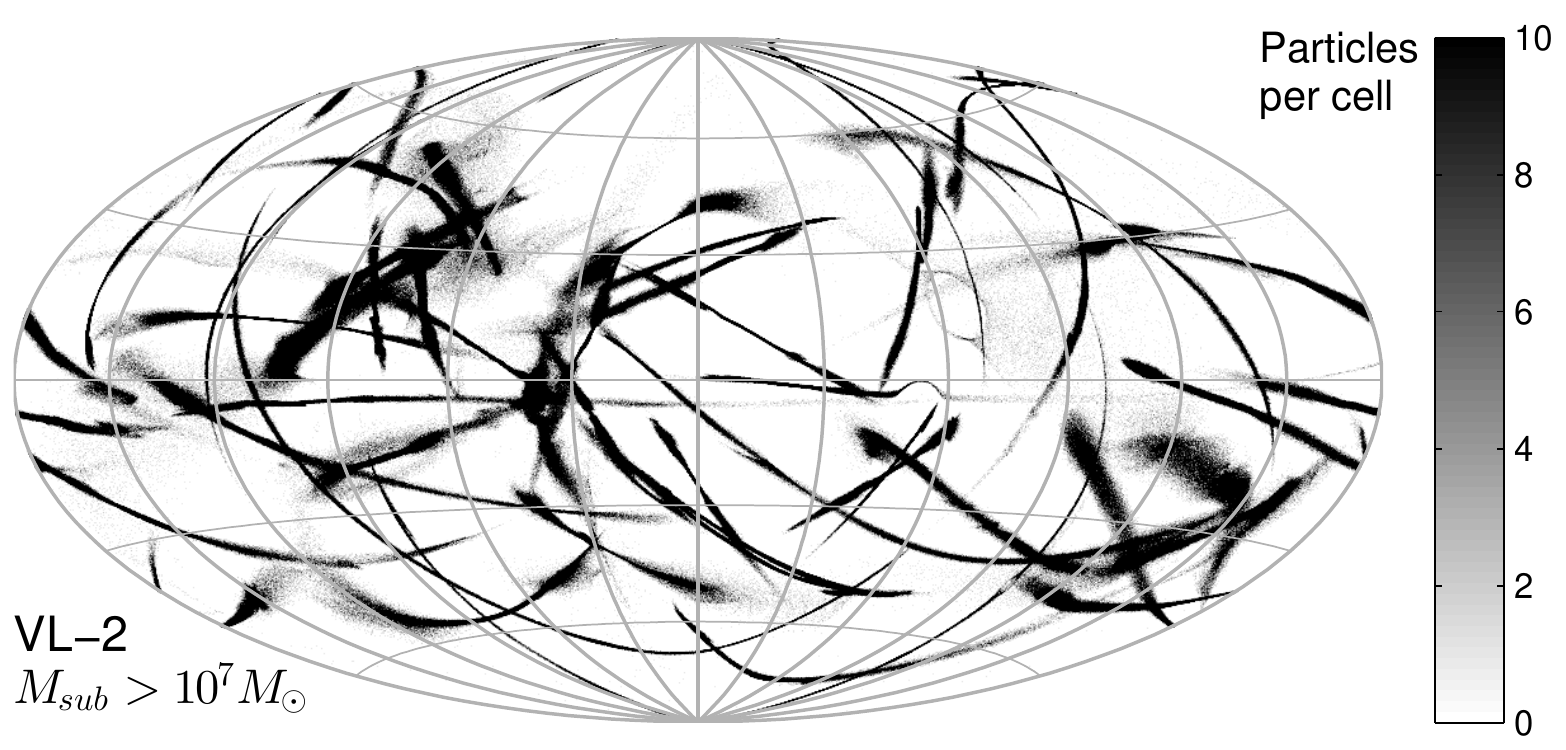}
	\hspace{0.05in}
	\includegraphics[width=3.4in]{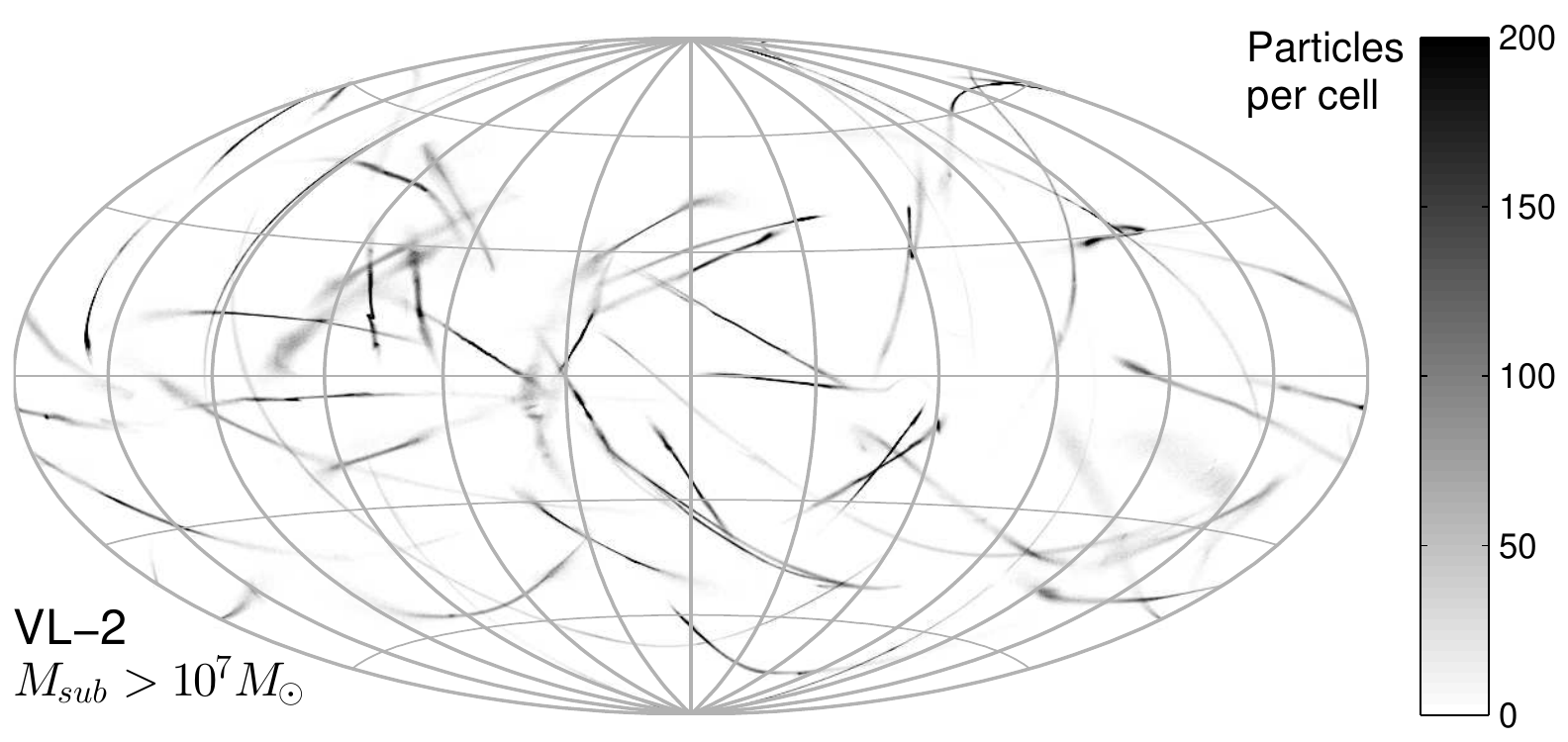}
	\caption{Similar to \fig{sky_NFW}, but the underlying halo is the VL-2 halo at redshift zero. Compared to the
	spherical halo in \fig{sky_NFW}, the streams in the VL-2 halo appear much more dispersed, especially in 
	smooth halo (top panels). \rev{Nevertheless, some streams in the lumpy halo can also be denser than the
	streams in the smooth halo (see \fig{count_errors}).} The orbits of the streams in this figure are selected by the same
	criteria and result in the same eccentricity distribution as the orbits in \fig{sky_NFW}.}
	\label{fig:sky_VLII}
\end{figure*}

\figs{sky_NFW}{sky_VLII} show the sky projections of 50 streams in each panel at 6 Gyr, as seen
from the galactic center, combined into the same images for 
the four cases of interest in this study. The top panels of the each figure show that the
{dynamical symmetry} of the smooth halo plays an important role in affecting the {dispersal} of the tidal debris. 
{A spherical potential has four isolating integrals of motion---energy and three components of angular momentum---which
constrain the phase space of the stream particles into only two dimensions
(\rev{for parts of the stream that are} far away from the progenitor). On the contrary, the lack of symmetry,
hence the {reduced number of} integrals of motion, in the VL-2 halo allows the stream particles to explore
the phase space with {fewer} constraints than when the potential is spherical. This is
why the stream particles in the VL-2 halo can have a wider variety of orbits, and the streams are
more diffuse than in a spherical halo.}

Comparing the top and bottom panels in each of {\figs{sky_NFW}{sky_VLII}} show 
that subhalos do not disperse streams as much as the lack of dynamical {symmetries} of the
underlying halos do. This finding is similar to the conclusion from \citet{SG08} which simulated 
the disruption of satellite galaxies that were much more massive and orbited at larger galactocentric
radii (ie. larger $s$ from Equation \ref{eqn:tidal_scale}) than our streams. In this
study we simulate the disruptions of a low mass globular cluster orbiting at smaller galactocentric radii, 
which result in streams that can be $\lesssim1\degrees$ wide as seen from the galactic center.

\rev{The right panels of \fig{sky_VLII} show another interesting result. Even though the streams in the VL-2 halos
are more diffuse than in the spherical halos, many of these streams remain very thin
in the VL-2 halos. This is true even in the presence of subhalos which serve as time-dependent fluctuations
for the streams. Therefore, thin globular cluster streams on moderately eccentric orbits (up to $e\leq0.5$)
are reasonably robust against the lack of 
dynamical symmetry in the halo, and stellar surveys in the future will likely uncover more thin streams if they exist.
}

\rev{\figs{sky_NFW}{sky_VLII} shows the progress over the past decade or so in 
simulating the influence that CDM subhalos have on globular cluster streams. Pioneering studies such as
\citet{ibata2002} and \citet{yoon11} simulated streams inside spherical halos (though the former included
a disk) in order to investigate how subhalos heat up and create gaps in streams, respectively.
In addition to subsequent studies such as \citet{carlberg12, carlberg13, NC14, erkal2015} which
use spherical halos, the streams in the above studies would be analogous to the ones shown in 
\fig{sky_NFW} here. In \citet{ibata2002} they also used a flattened but still idealized potential, and the
resulting streams would be an intermediate between \figs{sky_NFW}{sky_VLII}.
It was only recently when \citet{bonaca2014} and \citet{N14} simulated streams inside
the potential of the high-resolution halo of VL-2 directly, without fitting the potential to any idealized
profiles. Their results would be analogous to the streams shown in \fig{sky_VLII}. It is also worth
mentioning that \citet{johnston2002, mayer2002, SG08} also simulated streams with and without subhalos,
in a variety of profiles and shapes for the halo potential, but their streams were meant to be debris
from dwarf galaxies and were much wider than the streams shown in \figs{sky_NFW}{sky_VLII}.}


\subsection{{Dispersal} of Tidal Debris}
\label{sec:dispersion_of_tidal_debris}


\begin{figure*}
	\centering
	\includegraphics[width=7in]{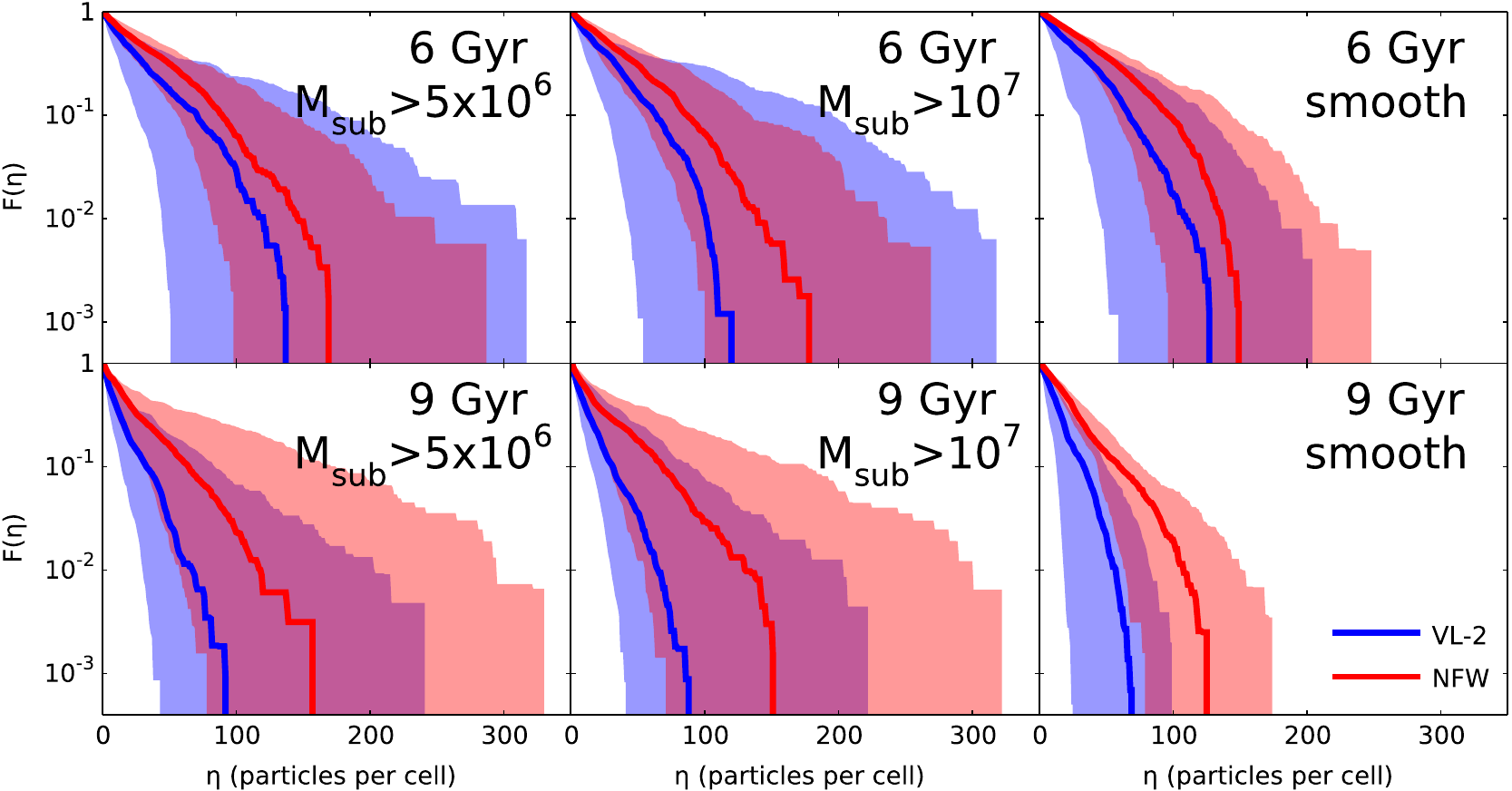}
	\caption{\rev{Cumulative fractions of the streams as functions of cell occupancy ($\eta$) in the VL-2 and the spherical
	NFW halos. The solid lines are the medians of 50 streams in each case. The colored areas enclose the points between
	the 15th and the 85th percentiles of the distribution ($\sim 1\sigma$). Top and bottom rows show the distributions
	at 6 and 9 Gyr, respectively. Left, middle, and right panels show the mass range of subhalos $M_{sub}$ present in the
	simulations, as labelled in units of solar masses.}}
	\label{fig:count_errors}
\end{figure*}

We now present an empirical analysis of the streams in the four cases of halos of interest.
For each stream, each particle is assigned into its nearest rectangular grid cell by its three-dimensional position.
The occupancies $\eta$ of the grid cells are similar to the number densities of the streams, but we caution
that our grid cells with 0.1 kpc on each side may be too coarse to be interpreted as measurements of density. This grid size 
is similar to the {transverse full-width-half-maximum} width of GD-1 at 70 pc wide
\citep{grillmair2006,gd1}, which is considered a very narrow stream with a {derived} orbit
\citep[peri- and apocenters at 14 and 29 kpc;][]{willett09} similar to orbits for our set of streams in this study.
Our grid size is chosen so that it is sensitive to 
the particles that have been dispersed away from the main track of the stream by more than that distance.

Note that even though the densities
of streams with different orbits cannot be compared directly, distributions of the cell occupancies $f(\eta)$ can tell us
how dispersed each stream is. If a stream remains narrow
without much {dispersal}, the occupancy of the grid would be dominated by a few densely occupied cells.
On the contrary, if a stream is very dispersed, then the occupancy of the grid would be spread among
many sparsely occupied cells.

\fig{count_errors} shows the medians and the spreads of the cumulative distributions of cell occupancies
of 50 streams in each of our halo potentials. The cumulative distributions have been weighted by the occupancies and then normalized
by the total numbers of particles. More precisely, the y-axes are
\begin{equation}
	F(\eta) = \frac{1}{N}\sum_{k>\eta}kf(k)
\end{equation}
where $N=50000$ is the total number of particles in each stream. \rev{So, $F(\eta)$ can be thought of as 
the fraction of the stream such that the cell occupancy is higher than $\eta$.}

\rev{At early times when the streams are $\lesssim5$ Gyr old, due to their moderate orbital eccentricities
the streams did not have time to develop extended tails. Furthermore, \fig{massloss} shows that at 
$\lesssim5$ Gyr some stream progenitors are still on the verge of being completely dissolved. For these streams,
their occupancy distributions would be dominated by the final remnants of the progenitors, which are trivially the
densest points in each stream. For these reasons, in \fig{count_errors} we show only the occupancy
distributions at later times such as 6 and 9 Gyr.
}

\rev{\fig{count_errors} is essentially the quantification of \figs{sky_NFW}{sky_VLII}. Recall that those two
figures clearly show that the streams in the VL-2 halo are much more diffuse than the streams in the spherical halo,
with or without subhalos in each case. This is quantified in \fig{count_errors}, where the streams in the VL-2 halo
(blue line and regions) consistently have lower $\eta$ than the streams in the spherical halo
(red line and regions) in every panel. At the diffuse end ($\eta\lesssim 100$ particles per cell), the fractions of
the streams can differ by as much as two orders of magnitude between streams in the two types of halos. As
discussed above, this can be attributed to the wider variety of orbits that the VL-2 halo permits due to its
lack of dynamical symmetry. In particular, extremely ``fluffy'' streams can be found in \fig{sky_VLII} and 
have been investigated in more detail by \citet{N14,pearson2014,pricewhelan2016}.
}


\subsection{Effects of subhalos}
\label{sec:effects_of_subhalos}

\rev{The tools developed in the previous section allow us to quantify how subhalos disperse tidal debris, which
is difficult to see by eye in \figs{sky_NFW}{sky_VLII}. The effects of subhalos are shown in the left and middle
columns of \fig{count_errors}. First of all, those two columns have negligible differences.
This means that subhalos with masses below $10^7\Msun$ do not contribute
significantly to dispersing tidal debris. Indeed, versions of \figs{sky_NFW}{sky_VLII} with $M_{sub}>5\sci{6}\Msun$
(not shown here) look almost identical to \figs{sky_NFW}{sky_VLII} with $M_{sub}>10^7\Msun$.
For the rest of our results, the ``lumpy halo'' refers to the case of a halo with the subhalo mass range $M_{sub}>10^7\Msun$.
}

\rev{Nevertheless, subhalos of masses $M_{sub}<10^7\Msun$ do inject enough energy to disperse tidal debris and produce
gaps in tidal streams. However, as shown in \citet{yoon11}, the change in energy of the debris decrease rapidly with 
distance from the impact, and the rate of this decrease depends on the subhalo's mass. More massive subhalos can inject
energy at parts of the stream farther away from the impact, whereas less massive subhalos can only do so locally. In other words,
the ``gaps'' produced by $M_{sub}>10^7\Msun$ subhalos are big enough that they may affect large parts of the streams. Our results
here focus on the effects of these massive subhalos.
}

\rev{A surprising feature can be seen when we compare streams in the smooth and the lumpy halos of the same halo type
(ie. same colors between middle and right columns of \fig{count_errors}).
Intuitively, one may think that subhalos can scatter stars away from a stream's
path, and this would decrease the $\eta$ of the stream. However, upon closer inspection of \fig{count_errors},
some streams in lumpy halos can behave in the opposite manner. At the diffuse end, $F(\eta)$ for the streams in 
lumpy halos are no lower than those in smooth halos. At the dense end, $F(\eta)$ for the streams in lumpy halos
are consistently much higher than those in smooth halos. Therefore, while the reduction of dynamical symmetry in
the halo can make streams more diffuse, the lumpiness of the halo can make certain parts of streams denser.
In the next section, we investigate how the densities in these streams are enhanced.
}

\rev{Density enhancements due to the lumpiness of the halo has important implication for stellar surveys in
the near future. The streams that are denser have higher surface brightness, so they may be more easily
found in surveys. This means that if the \lcdm\ prediction of the lumpiness of the Milky Way's halo
is true, then globular cluster streams should be easier to detect. This is a very encouraging result because the detailed
structures along these streams, in turn, contain crucial information about the \lcdm\ prediction itself
(see references in Section \ref{sec:introduction}).
}

%

%

\subsection{The Densest Streams {in the VL-2 Halos}}
\label{sec:the_densest_streams_in_the_vl2_halos}

\begin{figure*}
	\includegraphics[width=2.3in]{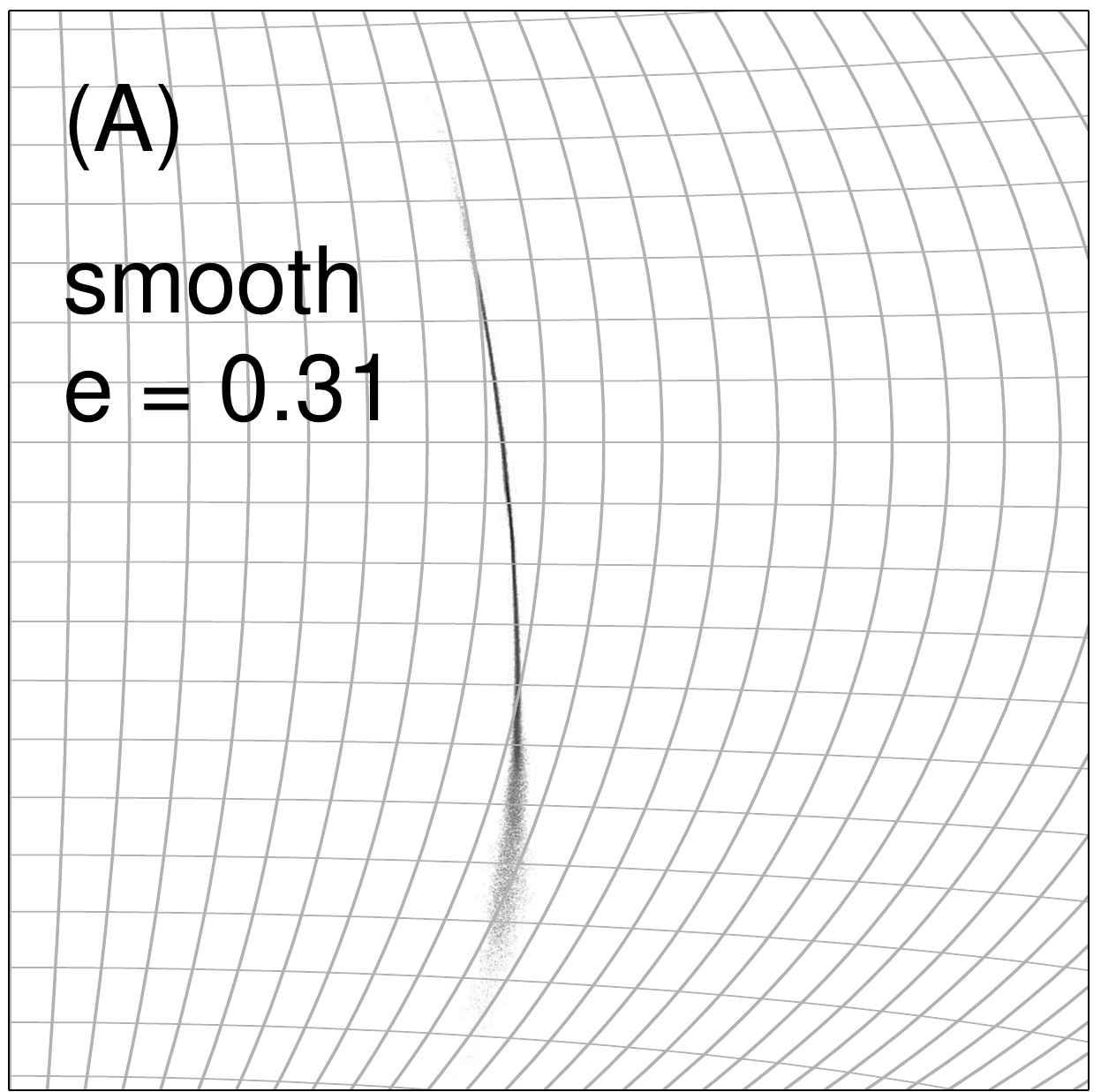}
	\includegraphics[width=2.3in]{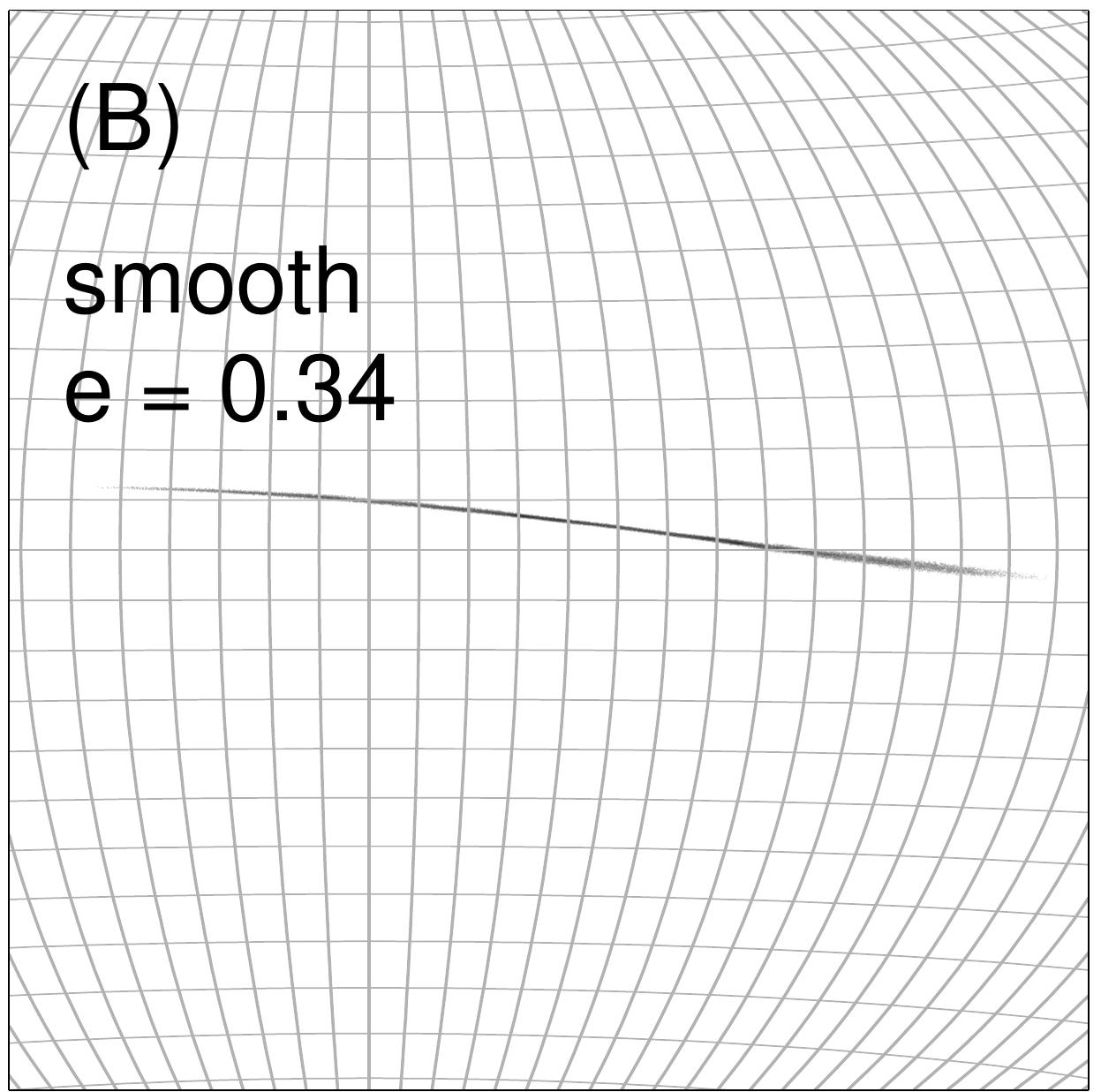}
	\includegraphics[width=2.3in]{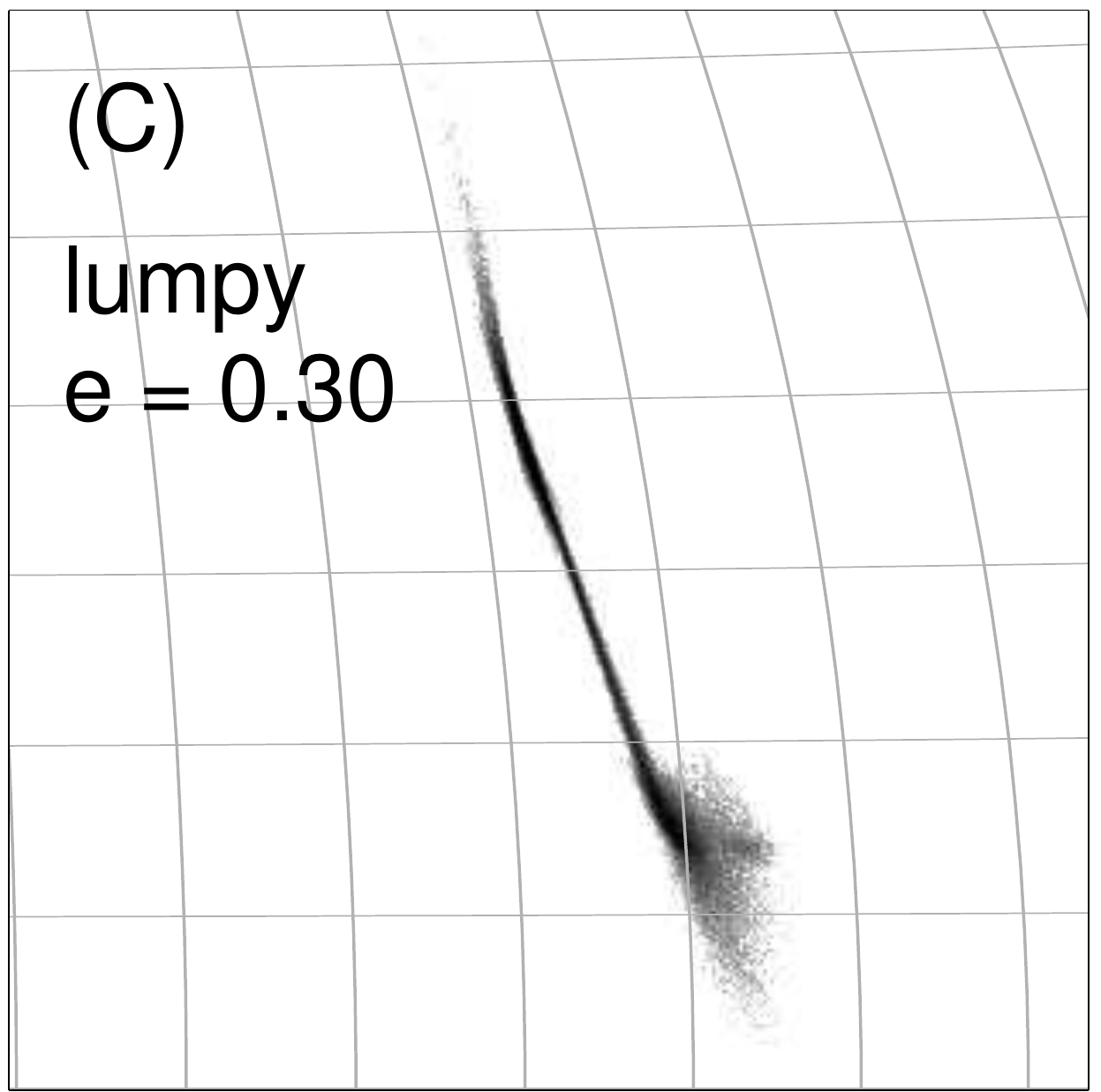}\\
	\includegraphics[width=2.3in]{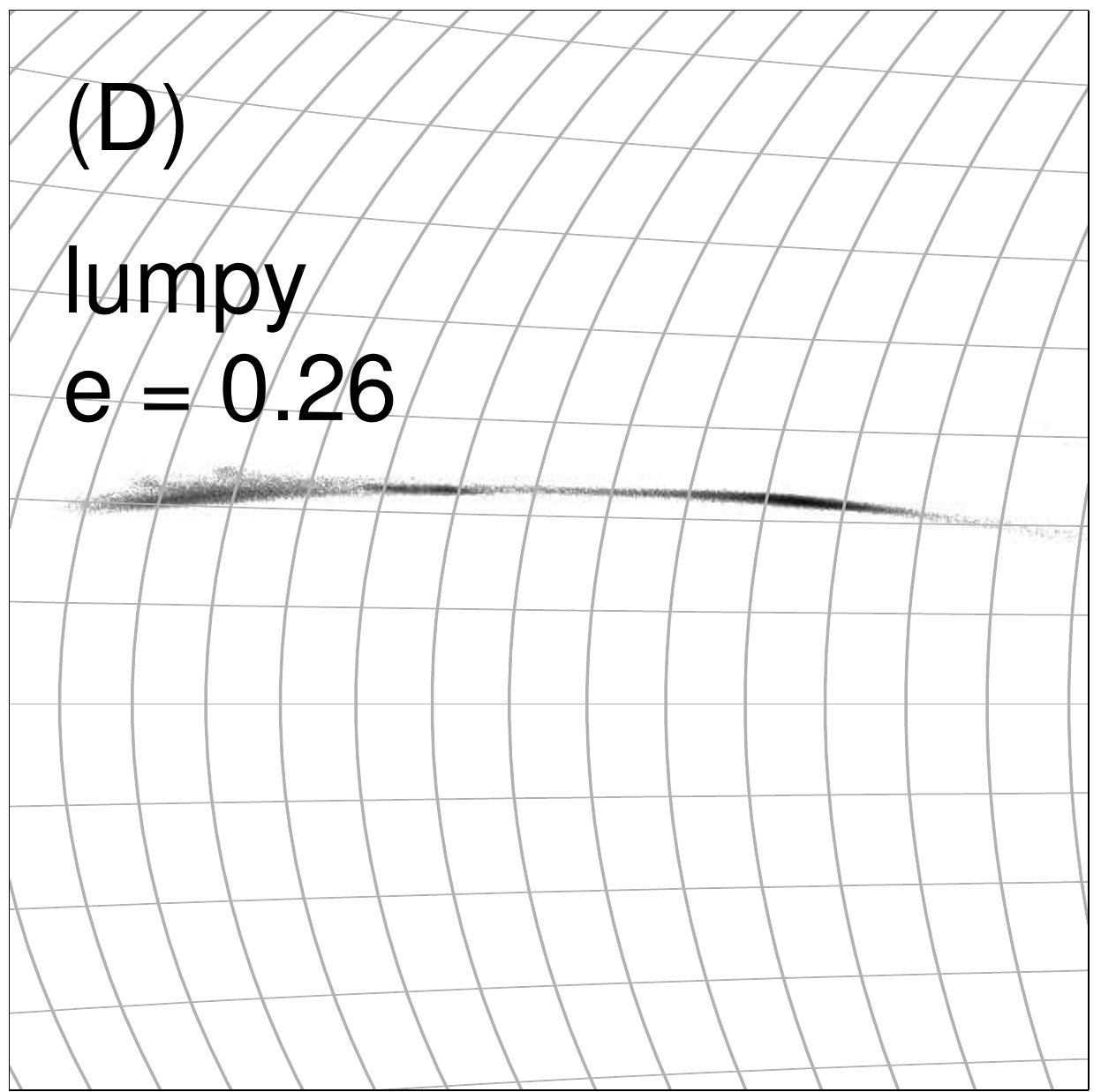} 
	\includegraphics[width=2.3in]{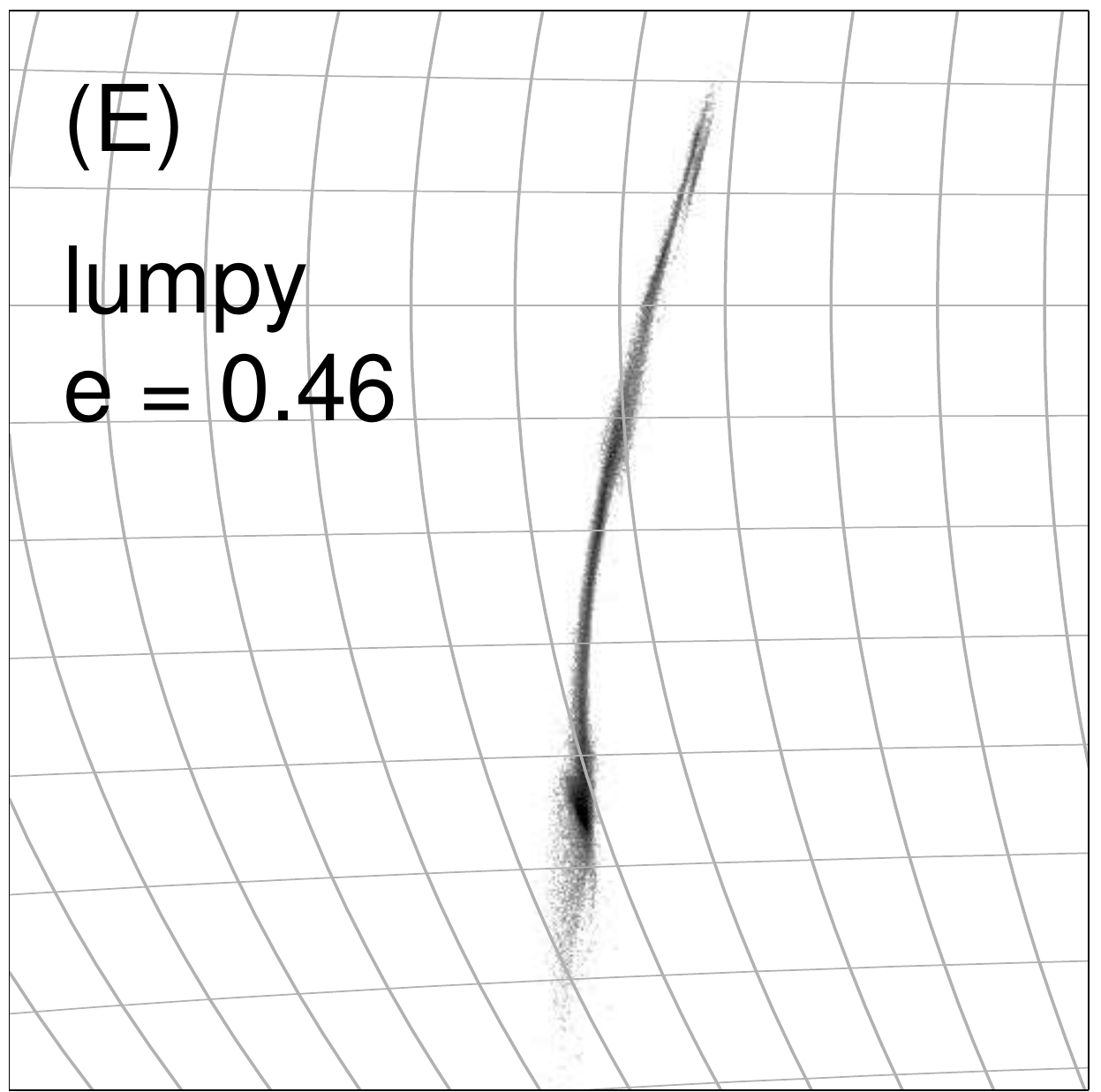}
	\includegraphics[width=2.3in]{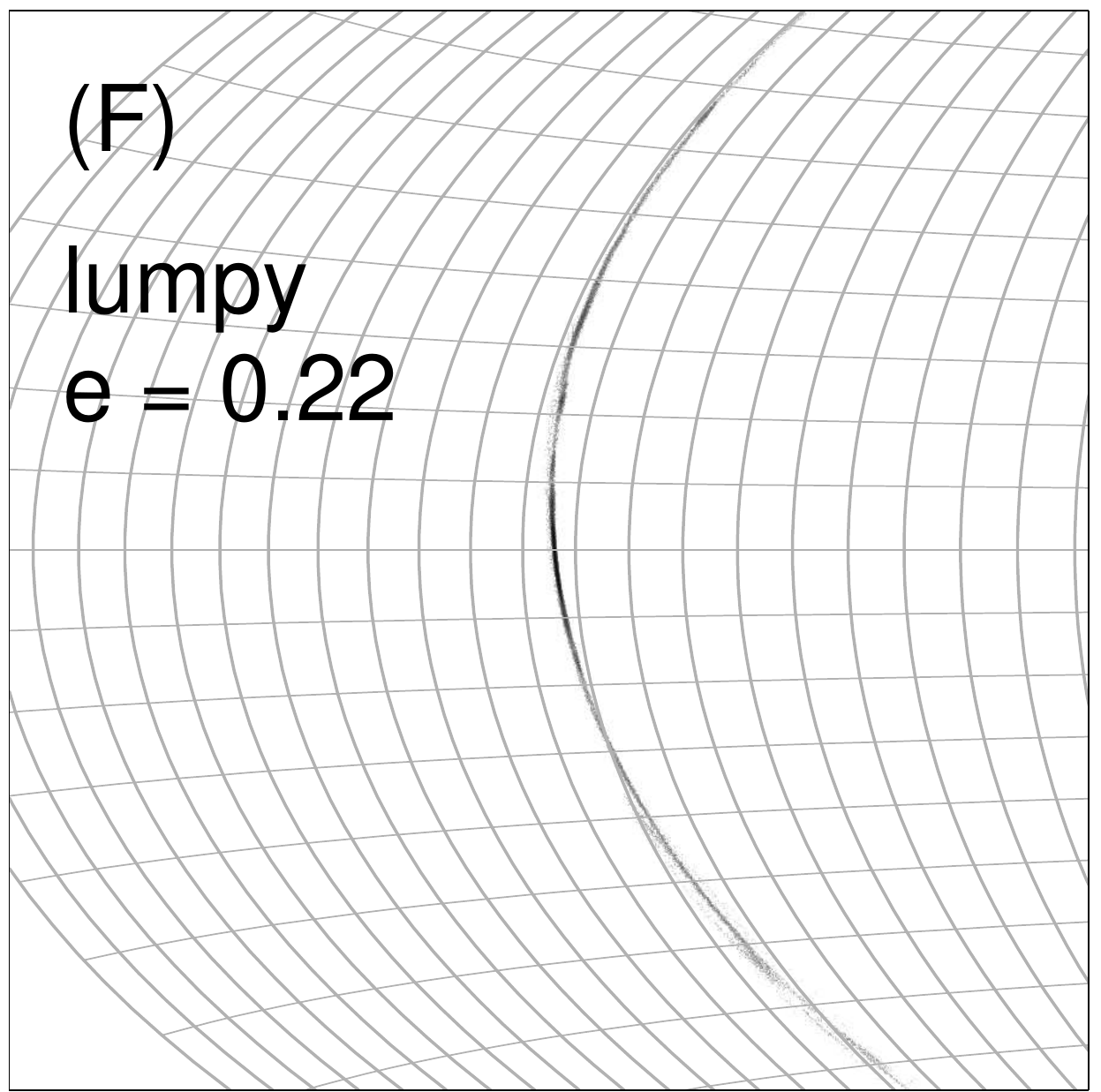}\\
	\centering
	\includegraphics[width=5in]{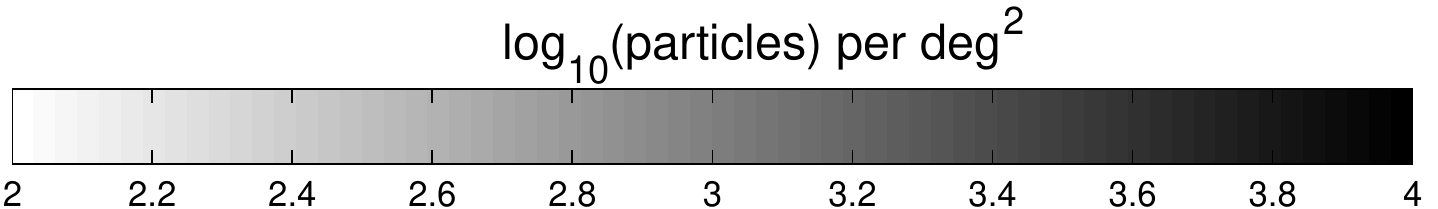}
	\caption{Six of the densest streams at 9 Gyr in the VL-2 halo. Each stream is projected onto the sky 
	using Hammer-Aitoff equal-area projection as seen from the galactic center. Both latitudinal and longitudinal
	grid lines in all panels are spaced at $5\degrees$ apart. \rev{The lumpy halo for streams C, D, E, and F contain
	subhalos at $M_{sub}>10^7\Msun$.} Eccentricities $e$ for the streams in the lumpy halo
	are quoted for the streams' orbits without subhalos. Note that the color scale is the log of the
	number of particles in each bin at $0.1\degrees$ on each side. In the smooth halo, stream A is undergoing a pericentric
	passage, but stream B is not. In the lumpy halo,
	\rev{the streams which contain the densest points are also thin, but the densest points are clumps that were 
	caused by interactions with subhalos.}}
	\label{fig:dense_streams}
\end{figure*}

\begin{figure}
	\centering
	\includegraphics[width=3.4in]{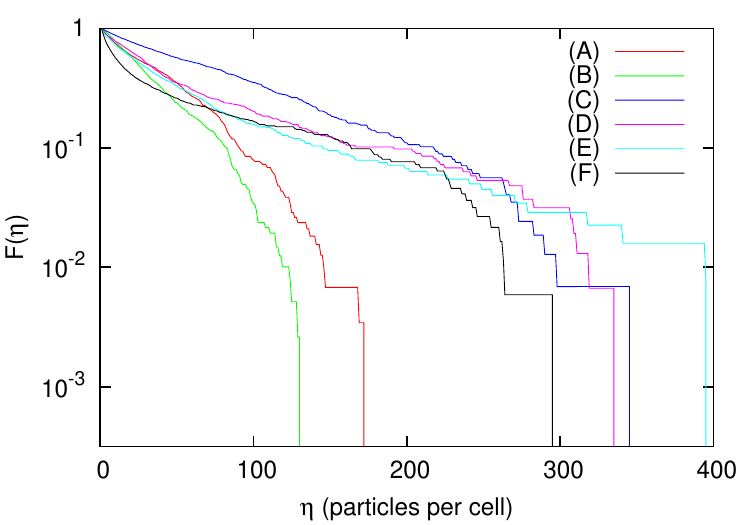}
	\caption{\rev{Cumulative distribution $F(\eta)$ of cell occupancy for the six streams shown in \fig{dense_streams}.
	Streams A and B are in the smooth VL-2 halo, and streams C, D, E, and F are in the lumpy VL-2 halo. These
	streams are outliers of the dense end in the distributions shown in blue in the bottom panels of \fig{count_errors}.}}
	\label{fig:dense_streams_plot}
\end{figure}


\rev{In this section we investigate the streams which orbit the VL-2 halo and contain the densest parts compared to other streams.
In order to avoid identifying the stream progenitors as trivially the densest parts of the streams, we only investigate
streams at 9 Gyr to ensure that all streams are completely dissolved (except one stream whose progenitor actually survived
for 10 Gyr. This stream is not being considered for the rest of this section). We follow our analysis in the previous section where
we assign the particles of each stream into a 3D grid with cell size of 0.1 kpc on each side. For each stream
we consider the cell with the maximum occupancy $\eta_{max}$, and we sort the streams by their $\eta_{max}$. 
The sky projections and $F(\eta)$ plots of six streams which contain the highest $\eta_{max}$ are shown in
\figs{dense_streams}{dense_streams_plot}. Note that these streams rank above the 85th percentile at the
densest end of the stream distributions, so the lines in \fig{dense_streams_plot} lie well outside of the blue regions in 
the bottom panels of \fig{count_errors}.

}



\rev{In the smooth VL-2 halo the densest streams are $\lesssim1\degrees$ thin as seen from the
galactic center. In \fig{dense_streams}
stream A is compressed transversely as it undergoes a pericentric passage, but stream B has just
gone past an apocentric passage.
This means that without subhalos, not only can thin globular streams survive in a realistic halo for
a Hubble time, but they are also the densest and easiest to find regardless of their orbital phases.
}


\rev{In the lumpy VL-2 halo, not only are the streams with the highest $\eta_{max}$ long and $\lesssim 1\degrees$ thin
similar to those in the smooth halo, but careful inspection of Streams C, D, E, and F in \fig{dense_streams} also reveals
density variations along them. Density variation in one stream in the VL-2 halo has previously been
reported in \citet{N14} but with a different subhalo mass range. Our study here shows that even in 
the presence of subhalos more massive than those ($10^8\Msun$) included in \citet{N14}, many streams remain thin
for a Hubble time.
}

\rev{The densest point of all the streams shown in \fig{dense_streams} is the clump at the lower tip of Stream E, whose
$\eta$ distribution is shown as the light blue line in \fig{dense_streams_plot}. That clump originated from a
direct impact by a subhalo with mass $M_{sub}=1.0\sci{8}\Msun$ and scale radius $r_s=0.8$ kpc at $\sim6.4$ Gyr.
The part of the stream which sustained the impact
was moving at $v_{stream}\simeq(107, -10, -120)$ \kms, and the subhalo was moving at $v_{sub}\simeq(143,378,-88)$ \kms.
Shortly after the impact at 6.46 Gyr (upper left panel of \fig{stream_E}), a density minimum can be seen at the point
of impact in between two density peaks. This is consistent with the matched filter profiles used in
\citet{carlberg12} and \citet{NC14} to look for gaps. The stream particles at the point of impact sustained changes
of energy of $\sim200\,\mathrm{km}^2\mathrm{s}^{-2}$, which agrees with the analytical estimate shown in
Figure 4 in \citet{yoon11}. At later times (upper right and lower left panels) the impact causes a shift in the 
stream particles' orbits, and the stream develops a ``z-fold'' which was seen in the idealized simulations in
\citet{carlberg09}. As the fold evolves with the stream, both tips of the fold can overlap and occupy the same grid cells.
This is seen in the lower right panel where the fold evolved into a clump by 8 Gyr, or 1.6 Gyr after the impact.
This is the reason that the typical $\eta_{max}$ for streams in the lumpy halo is $\sim2$ times higher than in the smooth halo.
}

\begin{figure}
	\centering
	\includegraphics[width=3.4in]{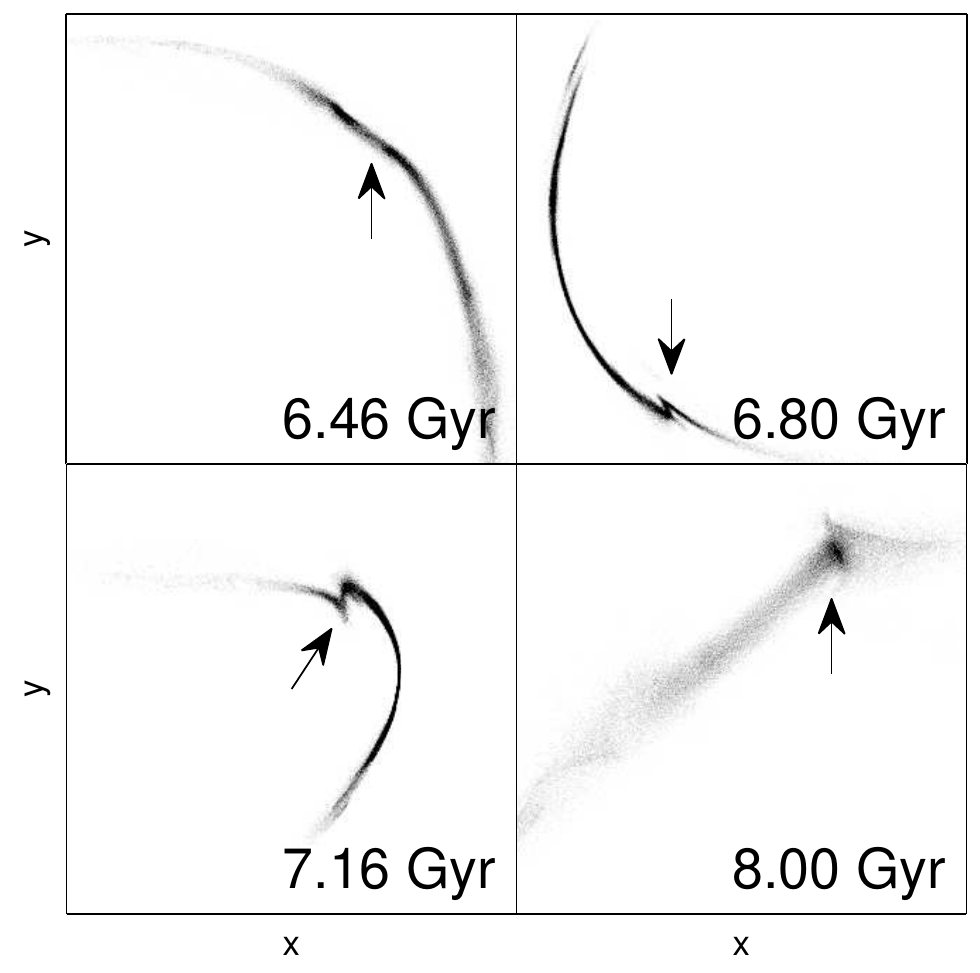}
	\caption{\rev{Time evolution of Stream E in \fig{dense_streams}. Each panel shows the surface
	density of the stream projected onto the $xy$-plane at different times as labelled. All panels are 16 kpc 
	on each side, and have the same color scale. The arrow indicates the position where a subhalo of
	$M_{sub}=1.0\sci{8}\Msun$ had a direct impact the stream at 6.42 Gyr, which later became a very dense clump
	along the stream.}}
	\label{fig:stream_E}
\end{figure}

\rev{All streams D, E, and F had encounters with $M_{sub}\gtrsim10^8\Msun$ subhalos where the impact parameter is less
than the scale radius of the subhalo. Stream C is an interesting case as it has only had one close encounter
with a $M_{sub}=1.6\sci{7}\Msun$ subhalo. The effect of this encounter is much less pronounced as that shown in
\fig{stream_E}, but it induces density variation along the stream nonetheless. Careful inspection of
\fig{dense_streams_plot} reveals that although Stream C does not have the highest $\eta_{max}$, at the intermediate
$\eta\lesssim250$ range it has the highest fraction compared to other streams. This is partially also due to its
apocentric approach where the stream is longitudinally compressed at that time. This compression, together with
the density variation induced by subhalos, makes Stream C one of the densest streams overall.
}

\section{Conclusion}

\label{sec:conclusion}

In this study we investigated the {dispersal} of \rev{a total of 300} tidal streams
which resulted from the disruption of a globular star cluster in four kinds of dark matter halos:
\begin{enumerate}[(a)]
	\item Spherical potential with no subhalos
	\item Spherical potential with orbiting subhalos
	\item Realistic potential with no subhalos
	\item Realistic potential with orbiting subhalos
\end{enumerate}
Both cases with subhalos (b and d) were further divided into two sub-cases with different subhalo mass ranges:
(i) all subhalos with $M_{sub}>10^7\Msun$, and (ii) all subhalos with $M_{sub}>5\sci{6}\Msun$.
In all cases the main halo was a time independent potential constructed using the 
zero redshift snapshot of the high-resolution dark matter halo in the Via Lactea II (VL-2) simulation.
The subhalos were extracted by a halo finder code from VL-2 and were also constructed as time independent
potentials, but they orbited around the smooth potentials.

For each case above, we simulated \rev{50} N-body streams whose progenitor orbits were inside galactocentric radii of
$10\,\mathrm{kpc} < r < 30\,\mathrm{kpc}$
in the smooth potentials (while allowing subhalos to scatter the tidal debris to arbitrary distances) for 10 Gyr.
\rev{The stream orbits were results from randomly distributed infall velocities, and the radial range were chosen 
to be rough matches with well-studied globular cluster streams such as Pal-5 and GD-1.}


For each stream we quantified the {dispersal} of the tidal debris by assigning all particles to their nearest
grid cells, and we plotted the distributions of cell occupancies. We found that the lack of {dynamical symmetry} of
the smooth halo \rev{can disperse} tidal debris more than subhalos do, similar to the conclusion
of \citet{SG08} which simulated the tidal disruptions of dwarf galaxies in larger orbits. \rev{On the other hand,}
we found that subhalos with $M_{sub}>10^7\Msun$ can make some streams much denser, hence 
more easily detectable, by bunching up stream stars into clumps which were denser than streams in the 
smooth halo. Meanwhile, subhalos with $M_{sub}>5\sci{6}\Msun$ produced almost the same distribution
of cell occupancies as the $M_{sub}>10^7\Msun$ case. Therefore, \rev{even though subhalos with masses
below $10^7\Msun$ can produce gaps in streams as shown in previous studies, these subhalos are not important
for globally redistributing the material in a stream.}

We selected a few streams which produced the highest cell occupancies\rev{---hence, most easily detectable---}in
the VL-2 halo in order to study their morphologies.
In the smooth halo, these dense streams are long and $\lesssim1\degrees$ thin as seen from
the galactic center. This suggests that even though a realistic halo disperses tidal debris, long and thin streams
can still survive for a Hubble time. In the lumpy halo, the streams with the highest cell occupancies were
just as long and thin as those in the smooth halo; however, the densest parts of the streams 
were primarily due to subhalo perturbations which caused stream particles to bunch up. This increased the occupancies
of the cells near the point where the bunch up occurred. Combined with the effect of longitudinal compression
at the streams' apocentric passages, streams in a lumpy halo can be much denser than those in a smooth halo.}

\rev{To put our results into context, the streams in our study were similar to (but not physical models of)
globular streams such as Pal-5 and GD-1. These streams have been well modeled and studied, and are excellent
sources of knowledge about the dark matter halo in our own galaxy. It would be very useful if more streams
similar to them are detected in the future, should they exist. The question we sought to answer was whether the
lack of dynamical symmetry of the halo potential or the lumpiness of the halo would erase these streams,
thus discouraging us from searching for more. As summarized in \figs{sky_VLII}{dense_streams}, our conclusion is
that even though streams inside realistic cases are more dispersed, many of them can survive as long and thin
structures over a Hubble time. In fact, the presence of subhalos may even make some streams easier to detect.
}

%



While our N-body streams {are} orbiting realistic halo potentials, our simulations are still missing
a few effects. In addition to a dark matter halo, the Milky Way also has a baryonic galaxy \rev{which was not
included in our study. Previous studies such as \citet{dehnen04} and \citet{brooks14} (and references therein)
found that the galactic disk can influence the mass loss rates of globular clusters and dwarf galaxies, respectively.
However, the effect of a galactic disk on the dispersal of tidal debris has yet to be studied in detail.} Furthermore,
other than the subhalos' orbits, all our potentials are static since we constructed them using only the
redshift zero snapshot. In the hierarchical structure formation model, satellite systems are continuously accreted and
merged into the main halo. Therefore, we expect both the main halo and the subhalos to be evolving in time.
Finally, we selected our stream orbits randomly in both
positions and velocities, which likely resulted in an unrealistic distribution of orbits. 
Current surveys such as Gaia \citep{perryman2001} will be very valuable in providing detailed information of kinematics of globular clusters
and streams in the Milky Way for future studies.

%
%
%
%

The authors thank the referee for a very detailed and constructive review.
R.W., A.S., and B.B. (at JHU) are supported by NSF grant OIA-1124403, and P.M. by OIA-1124452.
Computations were performed on the GPC supercomputer at the SciNet HPC Consortium, as well as
Sunnyvale at the Canadian Institute of Theoretical Astrophysics. SciNet is funded by:
the Canada Foundation for Innovation under the auspices of Compute Canada; the Government of Ontario;
Ontario Research Fund - Research Excellence; and the University of Toronto.

\bibliographystyle{apj}

\bibliography{gc}

\end{document}